\def\VersionAuthor{}
\ifdefined \VersionAuthor
	\newcommand{\AuthorVersion}[1]{#1}
	\newcommand{\FinalVersion}[1]{}
\else
	\newcommand{\AuthorVersion}[1]{}
	\newcommand{\FinalVersion}[1]{#1}
\fi

\ifdefined \VersionAuthor
	\PassOptionsToPackage{svgnames,table}{xcolor}
	\documentclass[]{article}
\else
	\PassOptionsToPackage{svgnames,table}{xcolor}
	\documentclass[acmsmall, review=true]{acmart}

	\usepackage[utf8]{inputenc}

	\usepackage[ruled]{algorithm2e}
	
	\SetAlFnt{\small}
	\SetAlCapFnt{\small}
	\SetAlCapNameFnt{\small}
	\SetAlCapHSkip{0pt}
	\IncMargin{-\parindent}

\fi

\ifdefined\VersionAuthor
	\usepackage[backend=biber,backref=true,style=alphabetic,url=false,doi=true,defernumbers=true,sorting=anyt,maxnames=99]{biblatex} %
	\DeclareSourcemap{
		\maps[datatype=bibtex]{
			\map[overwrite]{
				\step[fieldsource=toappear, match={true},
					fieldset=note, fieldvalue={To appear.~{}}, append]
			}
		}
	}

	\renewbibmacro*{doi+eprint+url}{%
		\iftoggle{bbx:doi}
			{\color{black!40}\footnotesize\printfield{doi}}
			{}%
		\newunit\newblock
		\iftoggle{bbx:eprint}
			{\usebibmacro{eprint}}
			{}%
		\newunit\newblock
		\iftoggle{bbx:url}
			{\usebibmacro{url+urldate}}
			{}%
	}

\fi
\usepackage{graphicx}
	\graphicspath{{./figures/}}
\AuthorVersion{
	\usepackage{amsmath, amssymb}
}
\usepackage{longtable}
\usepackage{url}
\usepackage{multirow}
\usepackage{multicol}

\usepackage{pdflscape}

\usepackage{paralist} %

\newenvironment{ienumerate}
	{\begin{inparaenum}[\itshape i\upshape)]}
	{\end{inparaenum}}

\newenvironment{oneenumerate}
	{\begin{inparaenum}[1)]}
	{\end{inparaenum}}

\usepackage{pgfplots}
\pgfplotsset{compat=1.9}

\ifdefined\VersionWithComments
	\usepackage[pagewise]{lineno} %
	\linenumbers
	
\fi

\ifdefined \VersionLong
	\newcommand{\LongVersion}[1]{#1}
	\newcommand{\ShortVersion}[1]{}
\else
	\newcommand{\LongVersion}[1]{}
	\newcommand{\ShortVersion}[1]{#1}
\fi

\ifdefined\VersionAuthor
	\usepackage[ruled,vlined,linesnumbered]{algorithm2e}

	\definecolor{USPNcobalt}{HTML}{293358}
	\definecolor{USPNocre}{HTML}{8b7d6d}
	\definecolor{USPNblanc}{HTML}{ffffff}
	\definecolor{USPNceruleen}{HTML}{354878}
	\definecolor{USPNsable}{HTML}{ad947e}

	\usepackage[
			pdfauthor={Etienne Andre et al.},%
			pdftitle={Formalizing UML State Machines for Automated Verification -- A Survey},
			breaklinks  = true,
			colorlinks  = true,
		\ifdefined \VersionWithComments
		\fi
			citecolor   = USPNsable,
			linkcolor   = USPNocre,
			urlcolor    = USPNceruleen,
		]{hyperref}

\fi

\usepackage[capitalise,english,nameinlink]{cleveref} %
\crefname{figure}{\text{Figure}}{\text{Figures}} %

\definecolor{colorHead}{rgb}{.5, .5, .5}
\definecolor{colorBad}{rgb}{1, .5, .5}
\definecolor{colorGood}{rgb}{.5, 1, .5}
\definecolor{colorPartially}{rgb}{1, 1, .5}
\definecolor{colorUnclear}{rgb}{1, .75, .5}
\definecolor{colorIndirectly}{rgb}{.75, 1, .5}
\newcommand{\no}{\cellcolor{colorBad}\ensuremath{\times}}
\newcommand{\noUnsure}{\cellcolor{colorBad}\ensuremath{\times} (?)}
\newcommand{\cellColorYes}{\cellcolor{colorGood}}
\newcommand{\yes}{\cellcolor{colorGood}\ensuremath{\surd}}
\newcommand{\yesUnsure}{\cellcolor{colorGood}\ensuremath{\surd} (?)}
\newcommand{\yesMono}{\cellcolor{colorGood}\ensuremath{\surd{}*}}
\newcommand{\partially}{\cellcolor{colorPartially}\ensuremath{\circ}}
\newcommand{\unclear}{\cellcolor{colorUnclear}?}
\newcommand{\indirectly}{\cellcolor{colorIndirectly}\ensuremath{(\surd)}}

\newcommand{\UMLelement}[1]{\textcolor{okcolor}{\texttt{#1}}}
\ifdefined\VersionWithComments
	\definecolor{okcolor}{rgb}{0, 0.3, 0.4}
\else
	\definecolor{okcolor}{rgb}{0, 0, 0}
\fi

\newcommand{\AnimUML}{\textcolor{okcolor}{AnimUML}}
\newcommand{\CPNAMI}{\textcolor{okcolor}{CPN-AMI}}
\newcommand{\CPNTools}{\textcolor{okcolor}{CPNtools}}
\newcommand{\cspsharp}{\textcolor{okcolor}{CSP$\sharp$}}
\newcommand{\EMI}{\textcolor{okcolor}{EMI-UML}}
\newcommand{\fdr}{\textcolor{okcolor}{FDR}}
\newcommand{\hugo}{\textcolor{okcolor}{\textsc{hugo}}}
\newcommand{\hugort}{\textcolor{okcolor}{\textsc{hugo}/RT}}
\newcommand{\jack}{\textcolor{okcolor}{JACK}}
\newcommand{\nusmv}{\textcolor{okcolor}{NuSMV}}
\newcommand{\pat}{\textcolor{okcolor}{PAT}}
\newcommand{\proco}{\textcolor{okcolor}{PROCO}}
\newcommand{\promela}{\textcolor{okcolor}{PROMELA}}
\newcommand{\smv}{\textcolor{okcolor}{SMV}}
\newcommand{\spin}{\textcolor{okcolor}{Spin}}
\newcommand{\tabu}{\textcolor{okcolor}{TABU}}
\newcommand{\uppaal}{\textcolor{okcolor}{\textsc{Uppaal}}}
\newcommand{\usmmc}{\textcolor{okcolor}{USMMC}}
\newcommand{\veriUML}{\textcolor{okcolor}{veriUML}}
\newcommand{\vuml}{\textcolor{okcolor}{vUML}}

\ifdefined\VersionAuthor
	\usepackage{xspace}
\fi

\newcommand\viz		{\textcolor{okcolor}{viz.,\@\xspace}}

\newcommand\wrt		{\textcolor{okcolor}{w.r.t.\@\xspace}}

\newcommand\eg		{\textcolor{okcolor}{\emph{e.g.},\@\xspace}}

\newcommand\ie		{\textcolor{okcolor}{\emph{i.e.},\@\xspace}}
\newcommand\resp	{\textcolor{okcolor}{resp.\@\xspace}}
\newcommand\etal	{\textcolor{okcolor}{\emph{\textcolor{okcolor}{et al.}}\@\xspace}}

\newcommand{\adhoc}{\textcolor{okcolor}{ad hoc}}

\AuthorVersion{%
	\makeatletter
	\def\orcidID#1{\,\smash{\href{https://orcid.org/#1}{\protect\raisebox{%
		+1.25pt%
	}{\protect\includegraphics{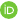}}}}}
	\makeatother
}

\newcommand{\ourAbstract}{%
	The Unified Modeling Language (UML) is a standard for modeling dynamic systems.
	UML behavioral state machines are used for modeling the dynamic behavior of object-oriented designs.
	The UML specification, maintained by the Object Management Group (OMG), is documented in natural language (in contrast to formal language).
	The inherent ambiguity of natural languages may introduce inconsistencies in the resulting state machine model.
	Formalizing UML state machine specification aims at solving the ambiguity problem and at providing a uniform view to software designers and developers.
	Such a formalization also aims at providing a foundation for automatic verification of UML state machine models, which can help to find software design vulnerabilities at an early stage and reduce the development cost.
	We provide here a comprehensive survey of existing work from~1997 to~2021 related to formalizing UML state machine semantics for the purpose of conducting model checking at the design stage.
}

\begin{document}

\ifdefined\VersionAuthor
	\sloppy

\else
	\acmJournal{CSUR}
	\setcopyright{acmlicensed}
	\acmJournal{CSUR}
	\acmYear{2023} \acmVolume{1} \acmNumber{1} \acmArticle{1} \acmMonth{1} \acmPrice{15.00}\acmDOI{10.1145/3579821}
\fi

\title{Formalizing UML State Machines for Automated Verification -- A Survey\AuthorVersion{\footnote{%
	This is the author version of the manuscript of the same name published in ACM Computing Surveys.
	The final version is available at \href{https://doi.org/10.1145/3579821}{\nolinkurl{10.1145/3579821}}.
}}}

\ifdefined\VersionAuthor
	\author{}
	\date{}
\else
	\author{\'Etienne Andr\'e%
	}
	\orcid{0000-0001-8473-9555}
	\affiliation{%
	\institution{Université de Lorraine, CNRS, Inria, LORIA}
	\streetaddress{BP 239}
	\postcode{54506}
	\city{Vandœuvre-lès-Nancy}
	\country{France}
	}
	\email{eandre93430@lipn13.fr}

	\author{Shuang Liu} %
	\orcid{0000-0001-8766-7235}
	\affiliation{%
	\institution{College of Intelligence and Computing, Tianjin University} %
	\city{Tianjin}
	\country{China}}
	\email{liushuangcs@gmail.com}

	\author{Yang Liu} %
	\orcid{0000-0001-7300-9215}
	\affiliation{%
	\institution{Nanyang Technological University}
	\streetaddress{50 Nanyang Avenue}
	\postcode{639798}
	\city{Singapore}
	\country{Singapore}
	}
	\email{yangliu@ntu.edu.sg}

	\author{Christine Choppy} %
	\orcid{0000-0001-7354-2925}
	\affiliation{%
	\institution{Universit\'e Sorbonne Paris Nord, LIPN, CNRS, UMR 7030, F-93430}
	\streetaddress{Avenue Jean-Baptiste Clément}
		\postcode{93430}
	\city{Villetaneuse}
	\country{France}}
	\email{christine.choppy@lipn.univ-paris13.fr}

	\author{Jun Sun} %
	\orcid{0000-0002-3545-1392}
	\affiliation{%
	\institution{Singapore Management University}
	\streetaddress{81 Victoria Street}
	\postcode{188065}
	\city{Singapore}
	\country{Singapore}}
	\email{junsun@smu.edu.sg}

	\author{Jin Song Dong} %
	\orcid{0000-0002-6512-8326}
	\affiliation{%
	\institution{National University of Singapore}
	\streetaddress{Computing Drive}
	\city{Singapore}
	\country{Singapore}
	}
	\email{dongjs@comp.nus.edu.sg}

\begin{abstract}
	\ourAbstract{}
\end{abstract}


\fi

\maketitle

\ifdefined\VersionAuthor
	\noindent{}\textbf{Étienne André\orcidID{0000-0001-8473-9555}}
	\\
	{\em\small{}Université de Lorraine, CNRS, Inria, LORIA, Nancy, France}

	\smallskip

	\noindent{}\textbf{Shuang Liu\orcidID{0000-0001-8766-7235}}
	\\
	{\em\small{}College of Intelligence and Computing, Tianjin University, China}

	\smallskip

	\noindent{}\textbf{Yang Liu\orcidID{0000-0001-7300-9215}}
	\\
	{\em\small{}Nanyang Technological University, Singapore}

	\smallskip

	\noindent{}\textbf{Christine Choppy\orcidID{0000-0001-7354-2925}}
	\\
	{\em\small{}Université Sorbonne Paris Nord, LIPN, CNRS UMR 7030, F-93430 Villetaneuse, France}

	\smallskip

	\noindent{}\textbf{Jun Sun\orcidID{0000-0002-3545-1392}}
	\\
	{\em\small{}School of Information Systems, Singapore Management University}

	\smallskip

	\noindent{}\textbf{Jin Song Dong\orcidID{0000-0002-6512-8326}}
	\\
	{\em\small{}National University of Singapore}

	\begin{abstract}
		\ourAbstract{}
	\end{abstract}

\fi

\section{Introduction}\label{sec:intro}

The Unified Modeling Language (UML)~\cite{UML251} is a standard for modeling dynamic systems.
UML behavioral state machines, an object-oriented variation of Harel's statecharts~\cite{Harel87}, can be
used to model the dynamic behaviors of a system.
The UML is considered to ``have become a \emph{de facto} `standard' for describing object-oriented design models, supported by a range of software tools and textbooks''~\cite{BBBKP11}.\label{newtext:defacto}
The UML specification, published and managed by 
the Object Management Group (OMG), is written in natural language.
Although this description is named ``formal specification'' and contains numerous details, it can be at most referred to as a ``semi-formal'' specification (a term often met in the literature).
This ``formalization'' in natural language introduces ambiguities and inconsistencies, which are tedious for manual detection or to be verified automatically due to the lack of formal semantics.
In fact, the meaning of ``formal'' in that ``formal specification'' could even be understood as ``formally adopted by the committee'', and not as in ``formal methods''.\label{newtext:formalspec}

Defining a formal semantics for UML state machines has been capturing a large attention since the mid-1990s.
The benefit of a formal UML state machine semantics is threefold.
Firstly, it allows more precise and efficient communication between engineers.
Secondly, it yields more consistent and rigorous models.
Lastly and most importantly, it enables automatic formal verification of UML state
machine models through techniques like model checking~\cite{BK08}, which enables the verification of properties in the early development stage of a system.
This results in a possible reduction in the overall cost of the software development cycle.

Since the mid-1990s, a number of works appeared in the literature,
which provide formalization for UML state machines usually for the purpose of performing model checking or more generally formal verification.
Those approaches adopt different semantic models, support different subsets of UML state machine features 
and only some of them are supported by an implementation.

\LongVersion{%
	We believe it is important to provide a comprehensive study and comparison on existing works which provide operational semantics for UML behavioral state machines, especially those that contribute to formal verification of UML state machines.
}%
In the past%
, three main surveys attempted to summarize the various methods for formalizing UML state machines, \viz{} \cite{BR04,CD05,LRS10}.
The work in~\cite{CD05} is the most complete and accurate, but it is now out-of-date, and hence does not cover recent approaches.
\LongVersion{%
	(We survey previous attempts to summarize the formalization of state machines in \cref{sec:related}.)
}%
		\cref{figure:timerange} (page~\pageref{figure:timerange}) shows that the formalization of UML state machines is still an active topic, and many recent works (typically all written after~2010) are not covered by any of these surveys.
\LongVersion{%
	In addition, no survey focuses specifically on formalizing UML state machines for the practical purpose of automated formal verification.
}

\paragraph{Contribution}
In this manuscript, we survey approaches aiming at formalizing UML behavioral state machines, with a specific focus on the automated verification support for UML state machines.
Our survey provides comparisons of those approaches in two dimensions.
The first dimension is the semantic model used by the approaches.
The existing approaches in the literature can be divided into two main categories:
\begin{enumerate}
	\item approaches that translate UML state machines into an existing formal language, and
	\item approaches that directly provide an \emph{ad-hoc} operational semantics for UML state machines.
\end{enumerate}
The approaches in the first category provide translation rules from UML state machines to some existing formal language,
such as 
\promela{} (which is the input language of the \spin{} model checker~\cite{SPIN04}) or extensions of automata and Petri nets.
The second category of approaches provides operational semantics for UML state machines, generally in terms of inference rules, which is a common format for formalizing structural operational semantics.

For the second dimension, we compare those approaches with respect to the covered features, such as entry or exit behaviors, run-to-completion step, deferred events, and tool support to automatically verify UML state machine models.

The contributions of the survey are as follows.
\begin{enumerate}
	\item Our first and main contribution is to provide an overview of the status of researches in the area of formalizing UML state machines.

	\item As a second contribution, we provide a study of tool supports for the formal verification of UML state machine models.
	We find out that, despite the interesting development of several formally grounded prototype tools, quite disappointingly most of these tools are unavailable nowadays, seemingly lost over the years.
	
	\item Lastly, we draw general conclusions, and identify future directions of research in this area.
\end{enumerate}

The rest of this survey is organized as follows.
\cref{sec:umlsm} briefly recalls the syntax and informal semantics of UML state machines. %
\cref{sec:cc} defines the categorization criteria of our survey.
\cref{sec:translation,sec:formalizing} discuss the translation approaches and the formalization approaches, respectively; we also summarize these works and draw some higher-level conclusions.
\cref{section:tools} surveys the existing tools supporting formal verification of UML state machines.
\cref{sec:related} discusses the related surveys. %
\cref{section:conclusion} concludes the paper and gives some perspectives.

\section{UML state machines}\label{sec:umlsm}

\paragraph{A brief history of UML versions}
After a draft (named ``1.0'') was proposed in 1996, the first actual version of the UML specification (named ``UML~1.1'') was proposed by the OMG in December~1997.
After a minor update named ``UML~1.2'' was released in July~1999, a more significant upgrade named UML~1.3 was released in February~2000.
Then, UML~1.4 was released in September~2001, followed by UML~1.5 in March~2003.

In July~2005, a major change was brought to the UML with the release of UML~2.0: several new diagrams were proposed, and several existing diagrams (including state machines) were modified.
Several subsequent updates followed:
UML~2.1.2 in October~2007,
UML~2.2 in January~2009,
UML~2.3 in May~2010,
UML~2.4.1 in July~2011
and
UML~2.5.1 in December~2017~\cite{UML251}.

In this section, we focus on the latest stable version, \ie{} UML~2.5.1~\cite{UML251}.
However, the various works surveyed here can address various versions of~UML, which we will discuss\LongVersion{ in the summarizing tables}.

\paragraph{UML state machines in a nutshell}
The \LongVersion{underlying }paradigm of UML behavioral state machines~\cite{UML251} is that of a finite-state automaton:
	that is, each entity (or subentity) is in one state at any time and can move to another state through a well-defined conditional transition.
We assume the reader's basic knowledge on state machines, and we only briefly recall here the syntax and informal semantics of state machines.
The UML provides an extended range of constructs for state machines:
simple/composite states, entry/exit/do behaviors, concurrency (regions in composite states, fork/join transitions), shared variables, shallow and deep history pseudostates, etc.
The use of composite states or submachine states allows to define state machines in a hierarchical manner.\label{newtext:hierarchy}
Communication is ensured via a (synchronous or asynchronous) broadcast mechanism, and through variables.

\begin{figure}[tb]
	{\centering
		\includegraphics[width=\textwidth]{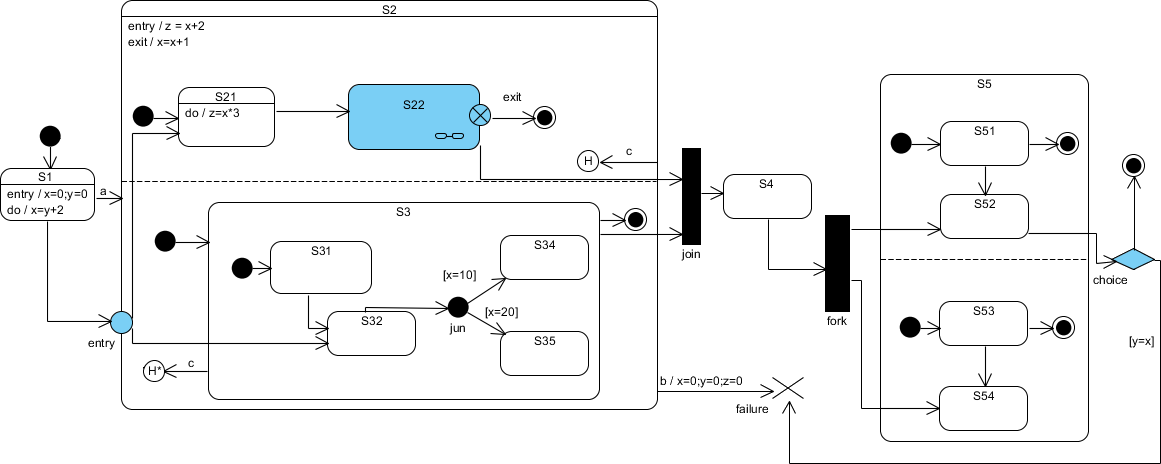}
	
	}
	\caption{An example of a UML state machine diagram}
	\label{fig:SMD}
\end{figure}

In the following, we briefly recall these elements.
We use as a running example the state machine diagram given in \cref{fig:SMD}.

\subsection{States}
The UML defines three main kinds of states:
	\emph{simple} states,
	\emph{composite} states (that can be \emph{orthogonal}),
and
	\emph{submachine} states.
A simple state (\eg{} \UMLelement{S1} or \UMLelement{S21} in \cref{fig:SMD}) has no region, and hence contains no ``internal state'' such as a region or submachine; internal transitions are still allowed in a simple state~\cite[p.\,320]{UML251}.

A \emph{composite state} (\eg{} \UMLelement{S2} or \UMLelement{S3} in \cref{fig:SMD}) is a state that contains at least one region and can be a \emph{simple} composite state or an \emph{orthogonal} state.
A simple composite state (\eg{} \UMLelement{S3} in \cref{fig:SMD}) has exactly one region, that can contain other states, allowing to construct hierarchical state machines.
An orthogonal state (\UMLelement{S2} and \UMLelement{S5} in \cref{fig:SMD}) has multiple regions (regions can contain other states), allowing to represent concurrency.
Regions are separated using a dashed line.

A \emph{submachine state} (\eg{} \UMLelement{S22} in \cref{fig:SMD}) refers to an entire state machine that can be nested within a state.
It is said to be ``semantically equivalent to a composite State'' \cite[p.\,311]{UML251}.
	(In \cref{fig:SMD}, the actual definition of \UMLelement{S22} is not given for sake of conciseness.)

\paragraph{Final states}
A final state (\eg{} the two right-most states in \UMLelement{S5} in \cref{fig:SMD}) is a special kind of state enclosed in a region to indicate when the region has completed.
(The UML assumes that the state machine itself is a region, which explains that a final state can be defined at level zero, \ie{} the final state right of \UMLelement{S5} in \cref{fig:SMD}.)

\subsection{Behaviors}\label{ss:behaviors}
Behaviors may be defined when entering states (``entry behavior''), when exiting states (``exit behavior''), while in states (``do behavior'') or when firing transitions.
\LongVersion{%

}%
The entry behavior is executed when the state is entered:
this can be the case of a default entry (``graphically, this is indicated by an incoming transition that terminates on the outside edge of the composite state''~\cite[p.\,310]{UML251}), an explicit entry, a shallow or deep history entry or an entry point entry.
\LongVersion{%

}%
Similarly, the exit behavior is executed when the state is exited.
The exit behavior of a source state is not executed in the case of a local or internal transition.
(We give in \cref{ss:RTC} more information on the actual sequence of entry and exit behaviors to be executed when firing a transition.)
\LongVersion{%

}%
The do behavior is executed only after the execution of the entry behavior of the state, and continues to be executing (in parallel with others behaviors, if any) until it completes or the state is exited.
\LongVersion{%

}%
In \cref{fig:SMD}, the entry behavior of \UMLelement{S2} is ``\texttt{z = x+2}'' while its exit behavior is ``\texttt{x = x+1}''.

\subsection{Pseudostates}
A main difference between a state and a pseudostate is that the active state configuration of a system only consists of \emph{states}---but not of pseudostates.
The UML defines different kinds of pseudostates\LongVersion{, described as follows}.

\paragraph{Initial}
	Each composite state (simple or orthogonal) and each state machine may have an initial pseudostate (\eg{} the left-most node in \cref{fig:SMD}).
\paragraph{History}
Only composite states can have, at most, one history pseudostate in each region.
	The UML defines two kinds of history pseudostate: \emph{shallow history} pseudostates and \emph{deep history} pseudostates.
	
	A \emph{shallow history pseudostate} is a kind of variable that represents the most recent active state configuration of its containing state, but not the substates of that substate, which means that the pseudostate saves only the latest visited state inside its containing composite state. 
	Shallow history pseudostates are depicted using an ``\UMLelement{H}''.
		In \cref{fig:SMD}, there is one shallow history pseudostate in the upper region of~\UMLelement{S2}.
	
	A \emph{deep history pseudostate} is a kind of variable that represents the most recent active state configuration of its owning state.
		That is, a deep history pseudostate saves the most recent active state configuration of all visited states inside the containing composite state (the state configuration is restored when a transition is terminating on the pseudostate).
	Deep history pseudostates are depicted using an ``\UMLelement{H*}''.
		In \cref{fig:SMD}, there is one deep history pseudostate in the lower region of~\UMLelement{S2}.
	Let us illustrate the difference between shallow and history pseudostate on this example.
	Assume the lower region of~\UMLelement{S3} is currently in~\UMLelement{S32};
	assume this deep history pseudostate is visited (via trigger ``\UMLelement{c}'').
	Then the new active state becomes (again)~\UMLelement{S32} (and entry behaviors are again executed).
	However, if this deep history pseudostate were replaced with a \emph{shallow} history pseudostate, the new active state would become~\UMLelement{S31}, as only the first level of hierarchy (\UMLelement{S3}) would be memorized.

\paragraph{Fork and join}
	The \emph{join} pseudostate ``serves as a common target vertex for two or more transitions originating from vertices in different orthogonal regions'' \cite[p.\,311]{UML251}.
	Therefore, join pseudostates can be seen as a synchronization function; they cannot have a guard or a trigger (event).
	The outgoing transition from a join pseudostate is executed only after the execution of all incoming transitions.

	The \emph{fork} pseudostate serves ``to split an incoming transition into two or more transitions terminating on vertices in orthogonal regions of a composite state'' \cite[p.\,311]{UML251}.
	Similarly to the join pseudostate, the transitions outgoing from a fork pseudostate cannot have a guard or a trigger. %
	
	For example, in \cref{fig:SMD}, the pseudostate source of the transition with \UMLelement{S4} as a target is a join pseudostate, while the pseudostate target of the transition with \UMLelement{S4} as a source is a fork pseudostate.
\paragraph{Choice and junction}
		The UML defines two kinds of pseudostates that allow to merge and/or choose between various flows, \ie{} they can have multiple incoming and outgoing transitions.
		The transitions can be guarded by Boolean expressions, and the system chooses nondeterministically one of the outgoing transitions for which the guard evaluates to true.
		The main difference between choice and junction pseudostates is that the guards are evaluated dynamically (after the exit behaviors are performed, and also after the effect of transition segments before reaching the choice) in choice pseudostates, whereas they are evaluated statically (before any compound transition containing this pseudostate is executed) in junction pseudostates.
		In \cref{fig:SMD}, the right-most diamond with one incoming arc (from~\UMLelement{S52}) and two outgoing arcs is a choice pseudostate; and the node labeled with \UMLelement{jun} within~\UMLelement{S3} in \cref{fig:SMD} is a junction.

\paragraph{Terminate}
		Entering a terminate pseudostate (\eg{} the pseudostate labeled with ``failure'' in \cref{fig:SMD}) implies that the execution of the state machine is terminated immediately, without performing any behavior.

\paragraph{Entry and exit points}
		``An entry point pseudostate represents an entry point for a state machine or a composite state that provides encapsulation of the insides of the state or state machine.''~\cite[p.\,317]{UML251} %
		The exit point is the dual concept of entry point.
		For example, the node in the lower left on the border of \UMLelement{S2} is an entry point; if it is entered, the system enters the regions of \UMLelement{S2} via \UMLelement{S21} and \UMLelement{S32}, without passing through the (regular) initial pseudostates.
		Conversely, the right-most node of \UMLelement{S22} is an exit point of this submachine state.
		Note that entry and exit points in composite states are very close semantically to forks and joins.

\subsection{Transitions}
The UML defines three kinds of transitions: external, local and internal.
\begin{itemize}
	\item External transitions are between two different vertices.
		This transition exits its source vertex, and the exit behavior of the incoming state is executed. %
	
	\item Local transitions can only be defined within a composite state, and are such that ``the transition does not exit its containing state (and, hence, the exit behavior of the containing state will not be executed)''~\cite[p.\,314]{UML251}.
	
	\item Internal transitions are special local self-transitions (with same source and target).
	 ``The state is never exited (and, thus, not re-entered), which means that no exit or entry behaviors are executed when this transition is executed.''~\cite[p.\,314]{UML251}

\end{itemize}
For example, the transition to the history pseudostate of the upper region of \UMLelement{S2} and labelled with event \UMLelement{c} in \cref{fig:SMD} is a local transition.

Transitions are executed as part of a more complex \emph{compound} transition that takes a state machine execution from one stable state configuration to another.
For example, in \cref{fig:SMD}, when in \UMLelement{S4}, a compound transition can be taken, that contains the incoming and the two outgoing transitions of the fork pseudostate.

We also emphasize so-called ``inter-level transitions'', which are a kind of transition that crosses the border of some composite state (called ``multi-level'' in \cite{Seifert08report}).
For example, the transition from \UMLelement{S22} to the \UMLelement{join} pseudostate in \cref{fig:SMD} is an inter-level transition.

Each kind of transition can have
	a guard (\eg{} \UMLelement{y=x} on the right-most transition in \cref{fig:SMD}),
	a trigger (\eg{} \UMLelement{a} on the left-most transition in \cref{fig:SMD}), which can be seen as an event,
	a behavior (\eg{} \UMLelement{x=0;y=0;z=0} on the left transition leading to the terminate pseudostate), and can be a completion transition (a transition without event) or a transition with event.
A completion transition (never labeled with an event) is a transition that is taken when a state finished its activity; in case of composite states, all regions must be in their final state.

Often, variables (integers, Booleans, etc.)\ can be used in state machines (notably in behaviors), and then tested in guards and updated along transitions.
This is also the case of the examples of the specification (\eg{} %
	\cite[Figs.~14.24 and 14.25]{UML251}).
In \cref{fig:SMD}, \UMLelement{x}, \UMLelement{y} and \UMLelement{z} are (integer-valued) variables.

\subsection{Deferred events}
A deferrable trigger allows to postpone the handling of the ``request'' event occurrence to subsequent states.

\subsection{Run-to-completion paradigm}\label{ss:RTC}
A central notion in UML state machines is the \emph{run-to-completion step}, which we briefly recall\LongVersion{ here}. %
Events are processed one by one, and only when the state machine is in a stable configuration.
That is, an event cannot be processed during the processing of another event (entry or exit behaviors, etc.).

More specifically, an event in the event pool is selected; then, among the list of enabled transitions, one or more transitions will be executed for firing.
In an orthogonal composite state, different compound transitions with the same event can be executed (in an undefined order) during the same step.
This set has to be maximal, and conflict-free.
A priority mechanism is defined in the UML specification to solve conflicts, \ie{} to decide which transition will fire when such a choice is to be made: without going into full details, this mechanism will generally give higher priority to transitions in substates.
That is, ``a transition originating from a substate has higher priority than a conflicting transition originating from any of its containing states''~\cite[p.\,317]{UML251}.
Note that even with the transition selection algorithm, non-determinism remains: it may happen that different (sets of) transition(s) could be selected: only one such set will be actually selected, and therefore, the execution is non-deterministic.
When performing model checking, this means that \emph{all} such sets of transitions must be explored in order to assess some formal property.

Once a set of transitions is chosen for a given event, the following actions are performed:
\begin{oneenumerate}
	\item the active state exit behavior is executed, followed by the exit behaviors of the containing states, up to a ``least common ancestor'' (\ie{} the ``innermost'' region that contains both the source and the target state concerned by the transition);
	\item the transition behavior is executed;
	\item the entry behaviors are executed, starting from the outermost state in the least common ancestor region that contains the target state, until the target state entry behavior itself.
\end{oneenumerate}

This whole notion is rather delicate, and its proper handling will be one of the key features we will survey in this manuscript.

\section{Methodology and categorization criteria}\label{sec:cc}
\subsection{Methodology}\label{ss:methodology}

In this paper, we survey works formalizing UML state machines for the purpose of automated verification.
We conducted a systematic literature review (SLR, see \eg{} \cite{BB06,KBBTBL09,WP13}), explained in the following.
We collected works using different methods:
\begin{ienumerate}%
	\item going through the former surveys~\cite{BR04,CD05,LRS10}, 
	\item performing search engine requests, typically \href{https://dblp.uni-trier.de/}{DBLP} and \href{https://scholar.google.com/}{Google Scholar},
	\item collecting citations to our own works (notably \cite{LLACSWD13}), and
	\item collecting systematically relevant citations from and to all the aforementioned works, until reaching a fixpoint.
\end{ienumerate}

\subsubsection{Criteria for inclusion}\label{sss:inclusion}
\label{newtext:inclusion}
The works we survey must address as main goal the formalization of UML behavioral state machines, with automated verification as the ultimate objective.
We therefore exclude purely theoretical approaches, except when they can be (potentially) used for formal automated verification.

\paragraph{Time range}
We focus on works published between December~1997 and the end of December~2021.
The start date of our considered time range (December~1997) is the publication of the first official version (named ``UML1.1'') by the~OMG.
The end date (end of December~2021) corresponds roughly to the time of revising our survey; as the research on formalizing UML state machines is still ongoing, we have to set up an end date for the collection of the surveyed works, which can seem arbitrary.

\subsubsection{Criteria for exclusion}\label{sss:exclusion}
We exclude publications formalizing diagrams prior to the OMG formalization; this can be the case of ``Harel's statecharts'' (\eg{} \cite{MLS97}).
In particular, any work prior to December~1997 is excluded by our survey.

We exclude publications strictly subsumed by others; this is generally the case of journal versions of conference papers (\eg{} \cite{CHS00} is subsumed by \cite{CGHS00}, \cite{SCH01} is subsumed by~\cite{SCH02}, etc.).

We exclude publications in which the formalization of state machines is really too shallow; this can be the case of publications formalizing other UML diagrams (\eg{} activity diagrams), with only a small focus on state machines.

We exclude publications related to formalisms close to (but different from) UML state machines.
We notably exclude significant extensions of UML, such as krtUML (for which an executable semantics is discussed in~\cite{DJPV05}), or SysML (partially formalized in, \eg{} \cite{JS15}).
The same applies to UML-RT.
Similarly, the Foundational UML (fUML)~\cite{fUML15}, which aims at providing an executable semantics for a subset of the UML syntax, is excluded from our work.
We however discuss fUML in \cref{section:conclusion}.

We tabulate the non-selected works in \cref{table:rejected}, with a reason for exclusion.
We discuss some of these excluded publications in the following.

In a series of works~\cite{AKY99,AGM00,AG04}, Alur \etal{} discuss several issues related to \emph{hierarchic reactive machines}, and notably semantic issues.
While some of these works were discussed in the former survey~\cite{BR04}, we did not integrate this line of works in our survey, because their semantics differs too much from that of UML state machines.
In~\cite{AG04}, it is explicitly mentioned that the ``mode'' (as a central component of their behavioral description) has several strong semantic differences with Statecharts and UML state machines:
such differences include notably the fact that
\begin{ienumerate}
	\item transitions can originate from and target entry/exit points only,
	\item a default exit always retains the history, and
	\item the priority between transitions differs.
\end{ienumerate}%
In that sense, the semantics of~\cite{AG04} is closer to \textsc{Room}~\cite{SGW94} than UML state machines.
In~\cite{AY01}, hierarchical state machines are discussed; they differ too significantly from UML state machines to be integrated to our survey.
In~\cite{AMY02}, hierarchical reactive modules~\cite{AG04} are considered.
Again, while sharing some similarities with UML state machines, the semantics of actual UML state machines is not discussed.

In~\cite{BLAHL02}, a version of finite state machines called hierarchical state/event machines (HSEMs) is discussed; as explicitly mentioned in~\cite{BLAHL02}, their semantics differ from UML state machines.

In~\cite{ATK01}, a formalization of class and state machine diagrams is given via an axiom system.
The syntactic features of UML state machines are very scarce (not even composite states nor final states), which rules out this work as a real formalization attempt.
An implementation in the HOL theorem proving system is however proposed.

The case of the UMC framework~\cite{BFGM11} is somehow borderline:
ter Beek \etal{} propose a UML-like framework for systems modeled using a collection of UML-like state machines.
Properties can be specified using the UML-oriented state-based and event-based logic UCTL.
While UMC is claimed to (partially) match the UML informal semantics (``UMC makes certain assumptions that, while compatible with the UML standard, are not necessarily the only possible choice''~\cite{BFGM11}), UMC is not exactly UML either.
The UMC documentation~\cite{Mazzanti09} does not really give a formal semantics to~UML, but rather explains ``how UMC can be used to generate system models according to the UML paradigm''.
Also, it is mentioned that it is expected that ``users directly use UMC for writing their model specification''.
On the other hand, the UMC documentation~\cite{Mazzanti09} still handles some rather subtle aspects of the UML semantics, such as completion transitions, and recursive or parallel operation calls.
In the end, we do not integrate the semantic framework in our survey, but we do mention the KandISTI/UMC tool suite in \cref{ss:tools}.
\label{newtext:BFGM11}

Finally note that we still used as much as possible the ``excluded publications'' in our search, \ie{} to gather further references until reaching a fixpoint.

\begin{table}
	\footnotesize
	\caption{Candidate articles not selected}
	\label{table:rejected}
	\scalebox{\AuthorVersion{.9}\FinalVersion{1}}{
	\begin{tabular}{c c l}
		\rowcolor{colorHead}
		Work & Authors & Reason for exclusion
		\\\hline
		\cite{PS91} & Pnueli and Shalev (1991) & out of time range
		\\\hline
		\cite{US94} & Uselton and Smolka (1994) & out of time range
		\\\hline
		\cite{HN96} & Harel and Naamad (1997) & out of time range
		\\\hline
		\cite{HG97} & Harel and Gery (1997) & out of time range
		\\\hline
		\cite{BHHKPRT97} & Breu \etal{} (1997) & out of time range
		\\\hline
		\cite{BF98} & Bruel and France (1998) & no technical details %
		\\\hline
		\cite{EBFLR98} & Evans \etal{} (1998) & no technical formalization
		\\\hline
		\cite{HP98} & Harel and Politi (1998) & not focusing on UML SMDs
		\\\hline
		\cite{LBC99} & Lüttgen \etal{} (1999) & not UML SMDs
		\\\hline
		\cite{AKY99,AGM00,AY01,AMY02,AG04} & Alur \etal{} (1999--2004) & not focusing on UML SMDs
		\\\hline
		\cite{CH00} & Clarke and Heinle (2000) & not UML SMDs %
		\\\hline
		\cite{CHS00} & Compton \etal{} (2000) & seems subsumed by \cite{CGHS00}
		\\\hline
		\cite{BS00} & Börger and Schmid (2000) & not specifically focusing on SMDs
		\\\hline
		\cite{ATK01} & Aoki \etal{} (2001) & too shallow
		\\\hline
		\cite{BCR01} & Börger \etal{} (2001) & only local aspects of formalization
		\\\hline
		\cite{SCH01} & Shen \etal{} (2001) & subsumed by~\cite{SCH02}
		\\\hline
		\cite{BCR02} & Börger \etal{} (2001) & only local aspects of formalization
		\\\hline
		\cite{BLAHL02} & Behrmann \etal{} (2002) & not focusing on UML SMDs
		\\\hline
		\cite{BDM02} & Bernardi \etal{} (2002) & relies on~\cite{MCBD02}
		\\\hline
		\cite{DJVP03} & Damm \etal{} (2003) & complements \cite{DJPV02}
		\\\hline
		\cite{BGOOS04} & Bozga \etal{} (2004) & no formal semantics
		\\\hline		
		\cite{OGO04} & Ober \etal{} (2004) & subsumed by~\cite{OGO06}
		\\\hline		
		\cite{FL05} & Fox and Luangsodsai (2005) & no formal semantics
		\\\hline		
		\cite{DJPV05} & Damm \etal{} (2005) & focusing on krtUML
		\\\hline
		\cite{FN05,FNP06,FN07} & Furfaro \etal{} (2005--2007) & not UML SMDs
		\\\hline
		\cite{KFBJHZAK05} & Kyas \etal{} (2005) & no details on SMDs
		\\\hline
		\cite{FBM06} & Fekih \etal{} (2006) & reverse translation from~B
		\\\hline
		\cite{TS06} & Truong and Souquières (2006) & no details on SMDs
		\\\hline
		\cite{vdB06} & von der Beeck (2006) & focusing on UML-RT
		\\\hline
		\cite{DJH08} & Dubrovin \etal{} (2008) & focusing on the symbolic step
		\\\hline
		\cite{TH08} & Thierry-Mieg and Hillah (2008) & no details on SMDs
		\\\hline
		\cite{MGT09} & Mekki \etal{} (2009) & pattern-based subset of UML SMDs
		\\\hline
		\cite{BFGM11} & ter Beek \etal{} (2011) & UMC rather than UML
		\\\hline
		\cite{Miao11} & Miao (2011) & no formal semantics
		\\\hline
		\cite{ZD12} & Zurowska and Dingel (2012) & focusing on UML-RT
		\\\hline
		\cite{ACHHLT13} & Androutsopoulos \etal{} (2013) & not UML SMDs
		\\\hline
		\cite{BCESEGM13} & Butler \etal{} (2013) & no details on SMDs
		\\\hline
		\cite{JS15} & Jacobs and Simpson (2015) & focusing on SysML
		\\\hline
		\cite{KMRG15} & Knapp \etal{} (2015) & mostly subsumed by~\cite{KM17}
		\\\hline
		\cite{AAC20} & Amtoft \etal{} (2020) & not focusing on UML SMDs
		\\\hline
		\cite{AlFedaghi20} & Al-Fedaghi (2020) & no formal semantics
		\\\hline
		\cite{HMC22} & Haga \etal{} (2022) & out of time range
		\\\hline
		\cite{RKR22} & Rosenberger \etal{} (2022) & out of time range
		\\\hline
	\end{tabular}
	}
	
\end{table}
\subsubsection{Summary}

We ended up selecting 61 works (to be surveyed in the following, and integrated to our comparison tables \cref{table:translation:features,table:translation:tools,table:semantics:features,table:semantics:tools}).

\subsection{Categorization}\label{ss:categorization}
Some of the surveyed works directly provide an operational semantics for UML state machines in the form of inference rule or SOS (structured operational semantics); hence, dedicated verification tools can be developed based on these semantics.
This is what we call the \emph{direct approach}.
Other works conduct an \emph{indirect approach}: they translate UML state machines into an existing specification language, which is usually supported as an input language by model checking tools. 

Based on the above observations, we categorized the surveyed approaches in two dimensions.
The first and main dimension is whether the approach is a direct or an indirect approach.
Hence, we split the main body of our survey into two main sections:
\begin{enumerate}
	\item \cref{sec:translation} surveys the indirect approaches, \ie{} the translation-based approaches;
	\item \cref{sec:formalizing} surveys the direct approaches, \ie{} the approaches that define an operational semantics for UML state machines.
\end{enumerate}

As a second dimension, we compare the surveyed works on the features supported, semantic models used, UML specification version, etc. 

We focus in particular on
approaches with tool support. %
These tools are mentioned throughout the survey, and are then specifically gathered in \cref{section:tools}.
A main outcome of our survey concerning tools is that, quite unfortunately, many tools are nowadays unavailable.

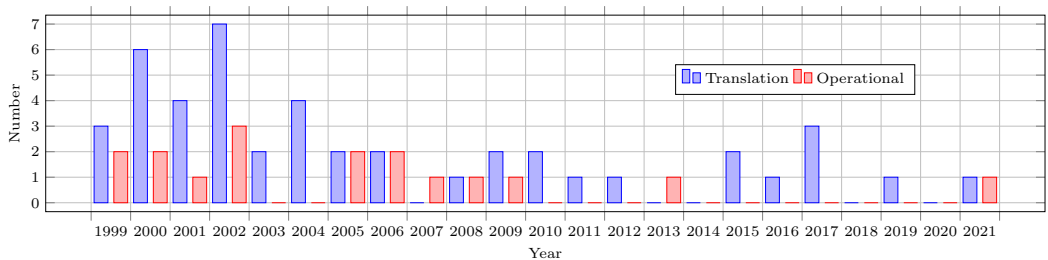
\begin{figure}[tb]
	\scriptsize
	\scalebox{.8}{
	\begin{tikzpicture}
		\begin{axis}[
			grid = major, %
			x tick label style	= { /pgf/number format/1000 sep=}, %
			ytick = {0,1,2,3,4,5,6,7,8},
			xlabel				= Year,
			ylabel				= Number,
			width				= \FinalVersion{1.2\textwidth}\AuthorVersion{1.5\textwidth},
			height				= .4\textwidth,
			enlargelimits		= 0.05,
			legend style		= {%
									at={(0.75,.75)},
									anchor=north,legend columns=-1},
			ybar interval		= 0.7, %
		]
		\addplot 
			coordinates {
				(1999, 3) %
				(2000, 6) %
				(2001, 4) %
				(2002, 7) %
				(2003, 2) %
				(2004, 4) %
				(2005, 2) %
				(2006, 2) %
				(2007, 0) 
				(2008, 1) %
				(2009, 2) %
				(2010, 2) %
				(2011, 1) %
				(2012, 1) %
				(2013, 0) 
				(2014, 0) 
				(2015, 2) %
				(2016, 1) %
				(2017, 3) %
				(2018, 0) 
				(2019, 1) %
				(2020, 0) %
				(2021, 1) %
				(2022, 0) %
			};
		\addplot 
			coordinates {
				(1999, 2) %
				(2000, 2) %
				(2001, 1) %
				(2002, 3) %
				(2003, 0) 
				(2004, 0) 
				(2005, 2) %
				(2006, 2) %
				(2007, 1) %
				(2008, 1) %
				(2009, 1) %
				(2010, 0)
				(2011, 0)
				(2012, 0) 
				(2013, 1) %
				(2014, 0) 
				(2015, 0)
				(2016, 0) 
				(2017, 0) 
				(2018, 0) 
				(2019, 0)
				(2020, 0)
				(2021, 1) %
				(2022, 0) %
			};
		\legend{Translation, Operational}
		\end{axis}
	\end{tikzpicture}
	}
	\caption{Number of selected works along the years}
	\label{figure:timerange}
\end{figure}

\paragraph{Time range}
We depict in \cref{figure:timerange} the number of publications per year of the surveyed works, by splitting them among the two identified categories.
While many works are concentrated around the 1999-2004 period, the research in formalizing UML state machines never really decreased, with still a number of works in the 2010s, up to the early~2020s.
And in fact, some very recent works even go beyond our time range (\eg{} \cite{HMC22,RKR22}), showing a still active field.

\section{The translation approaches}\label{sec:translation}

A popular approach to formalize UML state machines is to provide a translation to some existing  formal language (such as Abstract State Machines~\cite{ASM}, \ShortVersion{or }automata\LongVersion{, or Petri nets~\cite{Petri62} and their extensions such as colored Petri nets~\cite{JK09}}).
These formal languages have their own operational semantics, and they are usually the input languages of model checkers (such as \spin{}~\cite{SPIN04}, \nusmv{}~\cite{NuSMV}, or \uppaal{}~\cite{LPY97}).
This kind of approaches can be regarded as an indirect way of providing formalizations for UML state machines.
The purpose of this kind of approaches is to utilize existing verification techniques and tools. %

We categorize these approaches based on the target formal languages they adopt, \viz{}
	abstract state machines (\cref{ss:asm}),
	graph transformation (\cref{ss:graphtrans}),
	automata and extensions (\cref{ss:automata}),
	Petri nets and extensions (\cref{ss:PetriNets}),
	the input language of \smv{} and \nusmv{} (\cref{ss:smv}),
	extensions of CSP (\cref{ss:algebras}),
	PVS, KIV and (extensions of)~B
	(\cref{ss:PVSBZ}),
	and using institutions (\cref{ss:institutions}).
We summarize the translation-based approaches in \cref{ss:translation:summary}, and draw conclusions on these approaches.
\LongVersion{%

}%
In the following subsections, we only briefly mention the syntactic aspects covered by each work; the full list of these aspects will be summarized in \cref{table:translation:features}.

\subsection{Translation into abstract state machines}\label{ss:asm}
Abstract State Machines (ASMs)~\cite{ASM} offer a very general notion of state %
in the form of structures of arbitrary data and operations, which can be tailored to any desired level of abstraction.
State machines' configuration changes are represented by transition rules, which consist of conditions and update functions.
On the other hand, the notion of multi-agent (distributed) ASMs can naturally reflect the interaction between objects~\cite{ASM}.

\cite{Marc00,DGWtool00,JDSB08}
provide theoretic and tool support for model checking abstract state machines.

\LongVersion{%
We first review translation approaches to actual ASMs and their extensions (\cref{sss:asm}), and then to Object Mapping Automata, which are a variant of ASMs (\cref{sss:asm-variants}).
}

\subsubsection{Translation into ASMs}\label{sss:asm}

Börger \etal{}~\cite{BCR00,BRC03,BCR04} are among the pioneers in formalizing UML state machines using ASMs.
ASMs contain a collection of states and a collection of rules (conditional, update, Do-forall, etc.)\ which update those states. 

The first work in this direction is~\cite{BCR00} (in fact, this work relies on a previous attempt to formalize not state machines but activity diagrams \cite{BCR00act}).
This approach covers many UML state machine features, including 
shallow history and deep history pseudostates,
internal/external transitions,
deferred events, completion events and internal activities associated with states---which is important to note, as these features are discarded by most other approaches.
However, pseudostates such as fork, join, junction, choice (the latter being perhaps less straightforward) are surprisingly not directly considered.
The authors argue that these constructs can find their semantically equivalent constructs in their defined subset, where they use a  transition from (\resp{} to) the boundary of an orthogonal composite state to replace the join 
(\resp{} fork) pseudostates.

In 2003, another work~\cite{BRC03} extends~\cite{BCR00} to support transitions from and to orthogonal composite 
states\footnote{%
	The orthogonal composite state acts as the main source/target state of the transition, \ie{} the source/target of the transition can be a substate of the orthogonal composite state at any depth.
} in the context of event deferral and run-to-completion step.

In 2004, \cite{BCR04} provides some further discussions about the ambiguities in the official semantics of UML state machines and their solutions.

The works by Börger \etal{}~\cite{BCR00,BRC03,BCR04} cover a large set of features and the formalization is easy to follow due to the 
abstract feature of ASM notations.
However, to the best of our knowledge, no automatic translating tool was developed based on these works.

Another approach which also translates UML state machines into ASMs was proposed by 
Compton \etal{} in a technical report~\cite{CGHS00} (a shorter version was proposed in~\cite{CHS00}).
To be precise, this work translates UML state machines into \emph{extended} ASMs, 
in which ASMs are extended to represent inter-level transitions with multiple transitions 
which do not cross any boundary of states.
This extension makes it easier to deal with 
interruptions; it also makes the formalization 
procedure more structured and layered (since inter-level transitions break the hierarchical 
structure of UML state machine and such a decomposition of inter-level transitions into multiple 
transitions preserve such a hierarchical structure). 
Agents are used to process executions of UML state machines. An activity agent is used to 
model the execution of an activity associated with a node. 
The execution of agents is divided into different modes, which indicate what kind of rules (operations) the current agent should take.
Entry and exit behaviors are considered, as well as orthogonal states, guards, and completion events.
History pseudostates, deferred events\footnote{%
	In fact, the paper is a little ambiguous; on the one hand, it is explicitly stated that ``deferred events are not considered'' \cite[p.\,22]{CGHS00}; on the other hand, it seems that the ASM formalization does consider (at least in some definitions) deferred events.
}, forks, joins, choices, junctions are not considered.
Finally, the authors use \smv{} as a backend---which will be discussed in \cref{ss:smv}.

Finally, Jürjens also provides in~\cite{Jurjens02} a semantics in the form of abstract state machines, extending the semantics given in~\cite{BCR00}.
However, in contrast to~\cite{BCR00}, the focus of~\cite{Jurjens02} is not on supporting various features of UML state machine.
Instead, it rather focuses on the communications aspects between state machines.
The work explicitly models the message (with parameters) passing between state machines as well as the event queue.\footnote{%
	Note that the UML specification~\cite{UML251} mentions an event \emph{pool}.
}
Composite states, guards, events and (of course) message passing between various state machines is supported.
However, several common syntactic elements are discarded, \eg{} history pseudostates, transitions from or to composite states other than completion transitions, deferred events, fork/joins, choice/junctions, etc.
Note that consistency between various diagrams (including activity diagrams and class diagrams) is considered by the same author in~\cite{Jurjens02FMOODS}.

\subsubsection{Translation into Object Mapping Automata}\label{sss:asm-variants}

Jin \etal{}~\cite{JEJ04} provide an approach which syntactically defines UML statecharts 
as attributed graphs which are described using the Graph Type Definition Language (GTDL).
This work extends their previous work~\cite{JEJ02}.
They further provide some constraints in the form of predicates to specify the well-formedness 
rules of statecharts, which is considered as the static semantics of a UML statechart. 
The semantic domain is defined as an Object Mapping Automaton (OMA)~\cite{OMA98}, which is 
a variant of ASMs.
Given the abstract syntax (of the attributed graph) of a well-formed 
statechart, they first ``compile'' it into 
OMA algebraic structures, which specifies ``advanced static semantics'' of a UML statechart.
Based on OMA algebraic structures, two rules (\viz{} the initialization rule and the run-to-completion rule) are defined to describe the dynamic behavior of a UML statechart.
The syntax and semantics provided by this approach are more intuitive and easy to follow, benefiting from  the highly compatibility of the abstract syntax of attributed graph with UML statecharts.

This work is quite complete in terms of syntax, as its supports
completion events, history pseudostates (deep and shallow, including default entry), deferred events, entry/exit/do behaviors, internal and compound transitions as well as inter-level transitions.
\LongVersion{%
However, no tool support is provided.
}

\paragraph{Discussion}
Approaches translating UML state machines into ASMs tend to support more advanced features such as orthogonal composite states, completion/defer events, fork/join/history/choice pseudostates and inter-level transitions.
In \cref{table:translation:features}, the global score (last column), that is a way to measure the number of syntactic features considered, is among the highest of the translation approaches;
in particular, \cite{JEJ04} has the highest score of all approaches surveyed (together with~\cite{Traore00}, see \cref{ss:PVSBZ}).
A reason may be that ASMs are more flexible in terms of syntax format as well as update rules and are more suitable to express the non-structured feature of UML state machines.

\subsection{Translation via graph transformation}\label{ss:graphtrans}

In~\cite{Kuske01}, Kuske proposes a formalization of UML state machines using graph transformation.
The subset of syntactic elements is rather small, while not trivial either.
No automation of the translation is made, and no example of the verification of some property (even manually) is presented.
This approach is then extended with other UML diagrams in~\cite{KGKK02}.

In~\cite{KZDX09}, Kong \etal{} propose another approach to formalize UML state machines using graph transformation.
The supported syntax is much wider than in~\cite{Kuske01}; but, oddly enough, some features such as (entry/exit) behaviors do not seem to be supported.
It is also unclear whether the run-to-completion step is properly encoded.
This still makes this approach one of the most complete translation-based approaches.
\LongVersion{%
Finally, no implementation nor concrete verification is proposed.
}

\subsection{Translation into automata}\label{ss:automata}
\subsubsection{Translation into Extended Hierarchical Automata}
In~\cite{MLS97}, Mikk \etal{} propose extended hierarchical automata (EHAs) to encode the semantics of Harel's statecharts.
EHAs are basically extensions of finite-state automata with hierarchical capabilities very similar to the notion of hierarchy in state machines.
An operational semantics is given to EHAs, and a translation from statecharts to EHAs is then described.

Latella \etal{}~are among the first researchers who contributed to the formal verification of UML state machines.
\cite{LMM99} utilize EHAs defined in~\cite{MLS97} as an intermediate representation
of UML state machines; then they define the formal semantics of EHAs using Kripke structures.

Gnesi \etal{}~\cite{GLM99} propose a translation approach based on the formalization of UML state machines in their early work~\cite{LMM99}.
The translation is from a hierarchical automaton into a labeled transition system (LTS).
The LTS is then further translated into the FC2 format, which is the standard input format of \jack{}~\cite{JACKreport94}, an environment based on process algebras, automata and temporal logic formalisms.
The model checking is done \wrt{} a correctness property expressed in the action-based temporal logics ACTL.

\subsubsection{Translation into \promela{}}
\promela{} is the input language of the \spin{} model checker~\cite{SPIN04}.

Based on the formalization work using EHAs~\cite{LMM99}, Latella \etal{} proceed one step further in~\cite{LMM99Spin} by providing an automated translation from UML state machines to a \promela{} model.
The translation takes a hierarchical automaton as input and generates a \promela{} model as output.
A dedicated \promela{} process (called $\mathit{STEP}$) is defined to encode the run-to-completion step in UML state machines, which includes the following sequence:
\begin{oneenumerate}
	\item dispatching events from the environment;
	\item identifying candidate transitions to fire;
	\item solving conflicts and selecting firable transitions;
	\item performing the actual execution of the selected transitions (including identifying the next configuration after execution of the current transition and possible side effects, which are  events generated during the execution of actions associated with the transition).
\end{oneenumerate}
The run-to-completion step\LongVersion{ in a UML state machine} is, as indicated by the name itself, non-interruptable (but can be stopped\footnote{%
	The difference between interrupt and stop relies in the fact that interrupt means a temporary stop that needs to be resumed afterwards, whereas stop means a permanent stop without resuming.
}).
This non-interruptable nature is guaranteed by the \promela{} ``\textbf{atomic}'' command.
This mechanism guarantees in particular that ``the only values available for verification are those which variables evaluate to \emph{at the end} of each cycle''~\cite{LMM99Spin}.

The translation process is structured, since it is based on the predefined formal semantics of EHAs~\cite{LMM99}.
The authors also provide a proof for the translation to guarantee the correctness of the procedure \wrt{} the semantics they defined for UML state machines. 

Schäfer \etal{}~\cite{SKM01} provide a method to model checking UML state machines as well as collaborations 
with the other UML diagrams.
They translate on the one hand UML state machines into a \promela{} model, and on the other hand collaborations into sets of Büchi automata;
then invoke the \spin{} model checker to verify the model against  the automata.
Each state in the state machine is mapped to an individual \promela{} process. Two 
additional \promela{} processes are generated to handle event dispatching and transitions. The event 
queue is modeled as buffered channels and communication among processes are modeled via unbuffered 
channels, \ie{} they are synchronized. This approach further considers the consistencies between UML 
diagrams, \ie{} collaboration diagram and state machine diagram. The possible communications among 
objects shown in a collaboration diagram should be consistent with the dynamic behavior represented 
in the state machine diagram. By translating collaboration diagrams into sets of Büchi 
automata, which is the form of property to be checked against the model, this approach checks the consistencies between the two diagrams.
The approach is implemented in the tool \hugo{}.

Jussila \etal{} provide in~\cite{JDJLP06} another approach to translate UML state machines into \promela{} models. 
This approach considers multiple objects interacting with each other.
The translation is based on a formally defined semantics of UML state machines.
It supports initial and choice pseudostates as well as deferred and completion events. 
It further provides an action language, a subset of the Jumbala~\cite{Dubrovin06} action language, that is used to specify guard  constraints and the effects of transitions of a UML state machine.
The authors implemented a tool called \proco{}, that takes a UML model in the form of XMI files and outputs a \promela{} model.
Another translation for non-hierarchical state machines is also presented.

Carlsson and Johansson~\cite{CJ09} have designed a prototype tool to link \spin{} with RSARTE, a
modeling tool for UML diagrams. Their work focuses on all kinds of RT-UML diagrams, \ie{} UML diagrams 
related with real-time features. As part of UML, state machines are also translated into \promela{} 
in their approach. Since their work is not aiming at specifically model checking UML state machines, it does 
not provide detailed discussions about each feature of UML state machines, but discusses the 
communications between different objects.

\subsubsection{Translation into timed automata}

Timed automata are an extension of finite-state automata with \emph{clocks}, \ie{} real-valued variables that can be compared to integer constants along transitions (``guards'') or in discrete states (``invariants'')~\cite{AD94}.
They represent a powerful formalism to reason about systems featuring both concurrency and timing aspects.
Timed automata are supported by several model checkers, of which the most famous one is certainly \uppaal{}~\cite{LPY97}.

One of the earliest works using (an extension of) timed automata as the target formalism for formalizing UML state machines is performed by David \etal{} in~\cite{DMY02}.
A large subset of UML state machines is translated into \emph{hierarchical} timed automata, an \adhoc{} extension of timed automata.
These hierarchical timed automata are subsequently translated into ``flat'' timed automata.
Verification is done with \uppaal{}\LongVersion{, and properties checked include deadlock-freeness, safety and unbounded response properties}.

Knapp \etal{} present in~\cite{KMR02} another approach to translate timed UML state machines into timed automata.
Event queue and UML state machine are separately modeled by timed automata and the communication is modeled with a channel.
This approach is implemented in \hugort{}, which translates UML state machines into the \uppaal{} model checker, that can then verify whether scenarios specified by UML collaborations with time constraints are consistent with the corresponding set of timed UML state machines.
Note that the syntactic features displayed in \cref{table:translation:features} are those described in~\cite{KMR02}.
However, the tool \hugort{} was significantly enhanced since then, and all language constructs for UML state machines are now supported, with the exception of submachines, connection point references, and entry/exit points (see \cref{section:tools}).

Ober \etal{} present in~\cite{OGO04} an approach to translate a timed extension of a subset of UML state machines into communicating extended timed automata.
Not a lot of syntactic elements are considered, but two key concepts (the run-to-completion step, and proper priority handling) are encoded.
Some constructs seem to come from~\cite{MLS97}, but are not further detailed.
The IF toolset~\cite{BGOOS04} (whose formal language is based on communicating extended timed automata) is used as an underlying verification engine; properties can be specified using observer automata.
Then, Ober \etal{} define in~\cite{OGO06} (seemingly extending~\cite{OGO04}) a UML profile called ``OMEGA UML'', extending UML state machines using real-time extensions.
The syntax is translated into the input language of the IF toolset~\cite{BGOOS04}.
Again, properties can be specified using a lightweight extension of the UML syntax.
While this paper~\cite{OGO06} does not focus in detail on UML state machines (arguing they have been formalized in former works), again, the run-to-completion step is properly encoded.
Verification and simulation can be performed using the IF toolset.

Finally, Mekki \etal{} propose in~\cite{MGT09} an approach to translate UML statechart ``patterns'' into a network of timed automata.
The verification then reduces to reachability checking using a model checker for timed automata.
This approach is of particular interest, as it avoids the designer to understand all subtle aspects of UML state machines, and allows them to compose predefined patterns instead.
These patterns involve temporal and timed requirements (expressing concepts such as minimum and maximum delays, latency, simultaneity or sequence, in the line of other works relating patterns and timed automata~\cite{DHQS08,Andre13ICECCS}); concurrency (in the line of orthogonal composite states in UML state machines) is also considered.
We do not integrate this work in our subsequent summaries (\cref{table:translation:features}) though, as it does not take as input regular UML state machines, but only a set defined by an \adhoc{} ``patterns'' grammar.

\paragraph{Discussion}
From \cref{table:translation:features}, it is clear that the automata-based translation approaches do not support much of the syntax: no approach translating UML state machines into EHAs or \promela{} supports even half of the features.
However, a surprise is that \cite{KMR02,DMY02} are among the best of the approaches, with 9 (resp.\ 10)/17 features supported; this may come as a surprise as their works mainly focus on timed properties.
The run-to-completion step is notably properly encoded in~\cite{KMR02}, which is not often the case in translation approaches.

\subsection{Translation into Petri nets}\label{ss:PetriNets}

\LongVersion{%
Petri nets~\cite{Petri62} are a mathematical modeling language equipped with a graphical notation and (most importantly) with a formal semantics.
Petri nets aim at modeling distributed systems, with a focus on true concurrency.
}%
Petri nets~\cite{Petri62} are bipartite graphs (with places and transitions, and tokens evolving within places), and could be seen as an extension of automata;
still, we consider them in a separate subsection, in part due to the large literature translating UML state machines into (variants of) Petri nets.

Various extensions of Petri nets were defined.
In particular, colored Petri nets (CPNs)~\cite{JK09} are a special case of Petri nets in which the tokens are extended with attributes (types).
This results in a clearer and more compact representation.
Several approaches %
in the literature, notably in the 2010s, translate UML state machines into (possibly colored) Petri nets. 
We review them in the following.

Pettit~IV and Gomaa present in~\cite{PG00} an approach which uses  CPNs to model and validate the behavioral characteristics of concurrent object architectures modeled using UML.
UML collaboration diagrams are considered in particular.
The authors discuss how to map active/negative objects as well as message communications into CPNs.
Synchronous as well as asynchronous communications are discussed in  message communications.
Though not specifically dealing with UML state machines, this work provides a (very first) general idea of transforming UML diagrams to Petri nets. 
Verification is carried out using Design/CPN\footnote{\url{%
https://homepages.inf.ed.ac.uk/wadler/realworld/designcpn.html}}, which is an ancestor of \CPNTools{}~\cite{Westergaard13}.

Baresi and Pezzè propose in~\cite{BP01} another approach to formalize UML with high-level Petri nets, 
\ie{} Petri nets whose places can be refined to represent composite places.
Class diagrams, 
state diagrams and interaction diagrams are considered. Customization rules are provided 
for each diagram. 
However, the authors do not provide details about those customization rules; instead, they illustrate 
the steps with the ``hurried philosopher problem''.
The analysis and validation are also discussed, especially how to represent properties (such as absence of deadlocks or fairness) in UML, as well as how to translate them into Petri nets models.
Instead of providing an automatic tool, the paper discusses how model checking can be conducted on various properties by querying the existing CASE tools (taking high-level Petri nets as input).

In~\cite{SSH01}, Saldhana \etal{} propose an approach to formalize UML state machine diagrams using extensions of Petri nets.
First, state machine diagrams are translated into flat state machines; and then into OPNs (object Petri nets); then, UML collaboration diagrams are used to derive a colored Petri net.
The translation process is not very formal, but several examples illustrate it.
No experiments are performed, but the model of a Spacecraft Control System is thoroughly discussed.
This work is then extended to simulation in~\cite{HS04,LHS08}.

In~\cite{MCBD02}, Merseguer \etal{} translate state machines into generalized stochastic Petri nets %
(see \eg{} \cite{ABCDF98}).
Unfortunately, only ``flat'' state machines are considered, and therefore any hierarchical construct is disregarded, as well as pseudostates (except initial states).
No automated translation is proposed.
Then, in~\cite{BDM02}, Bernardi \etal{} translate state machines (and sequence diagrams) to generalized stochastic Petri nets.
The overall goal is to ensure consistency between the sequence diagrams and the statecharts.
The translation of state machines itself relies on the translation proposed in~\cite{MCBD02}.

In~\cite{TZ05,TZH05}, Trowitzsch and Zimmermann translate a subset of \emph{timed} UML state machines into stochastic Petri nets.
In~\cite{TZ05}, the authors use stochastic Petri nets, which contain exponential transitions, making it more suitable to model time events.
The approach covers a quite interesting subset of the UML state machine syntax, including time events.

In~\cite{CKZ11}, Choppy \etal{} propose an approach that formalizes UML state machines by translating them to colored Petri nets.
They provide a detailed pseudo algorithm for the formalization procedure.
They map simple states of UML state machine into Petri nets places and composite states of UML state machine into composite Petri nets places.
Transitions in UML state machines are mapped to arcs in Petri nets and corresponding triggering events are properly labeled. 
An extra place called $\mathit{Events}$ is modeled with an ``event place'' in Petri nets, in which each event type is translated into a different color.
Entry and exit actions of UML state machines are modeled with an arc in Petri nets which is labeled with the proper event type and ends in the event place.
Though the mapping from UML state machines to high-level Petri nets is clearly expressed compared to~\cite{BP01}, a very limited subset of UML state machine features is supported:
only the very basic features such as simple state, composite state, transitions, triggers and entry/exit actions are discussed.
How to deal with more complex concurrent composite states, and notably non-trivial forks or joins, is not discussed.
A translation prototype\footnote{%
	That we were not able to find online.
} has been implemented, and verification is carried out using \CPNAMI{}~\cite{HHKLPRT06}.

André \etal{} propose in~\cite{ACK12} an approach different from the work by~\cite{CKZ11}, and support a larger subset of UML state machine features, including state hierarchy (\ie{} composite states), internal/external transitions, entry/exit/do activities, history pseudostates, etc.
However, a limitation of that approach is that concurrency is left out: hence fork and join pseudostates, as well as
communication between state machines via triggers
is not considered.
Verification is carried out using \CPNTools{}~\cite{Westergaard13}.

\cite{ABC16} extends~\cite{ACK12} by reintroducing the concurrency; hence \cite{ABC16} supports the syntactic elements considered in~\cite{ACK12}, with the addition of fork and join pseudostates, orthogonal regions, as well as concurrent inter-level transitions.
The run-to-completion semantics is also properly handled.%
\footnote{%
	Note that some of the authors of this survey are involved in this series of works.
}

In~\cite{MCSK17}, Meghzili \etal{} propose another translation from UML state machine diagrams to colored Petri nets.
The subset of the syntax is rather limited (mostly orthogonal states, entry and exit behaviors).
However, a major feature is that the model translation is formalized in Isabelle/HOL, which makes it a \emph{verified} translation.
The verified transformation is then performed with Scala.
The authors extend their work in~\cite{MCSK19}, and a translation from BPMN (Business process) to CPNs is notably proposed to show the genericity of the approach.

In a parallel direction, Kumar \etal{} propose in~\cite{KST17}  safety analyses in terms of quantitative probabilistic hazard assessment; the authors convert (a quite restricted subset of) UML state machine diagrams into Petri Nets.
Probabilistic model checking is then used; a reactor core isolation cooling system is used as a case study.

In~\cite{LM19}, Lyazidi and Mouline use Petri nets to model the behavior of UML state machines.
Model checking is performed against safety properties, as well as deadlock-freeness and liveness.
A quite restricted subset of the UML syntax is considered, mostly simple states and fork/join pseudostates.
The transformation is automated, and the resulting model is given in the TiNA syntax~\cite{BV06}; it remains unclear whether timing aspects are actually considered (for which the use of TiNA would make sense).

\paragraph{Discussion}
Petri nets are used both in the academics, and the industry (in particular for modeling workflows).
But they may be more difficult to understand for non-experts than UML.
With automatic translators from UML state machines to Petri nets, engineers can benefit from the rigorous verification power of existing Petri nets verification tools.

However, approaches translating UML state machines to Petri nets usually cover a small subset of UML state machine features: with the exception of~\cite{ABC16}, no approach supports more than half of the UML syntax (see \cref{table:translation:features}); in addition, and again with the exception of~\cite{ABC16}, no Petri nets translation approach supports the notion of run-to-completion step.
Even the support of the run-to-completion step in~\cite{ABC16} was not easy to manage, and required a separate encoding.

This little syntactic support of approaches based on translations to Petri nets for formalizing UML state machines can be seen as a paradox for two main reasons.
First, Petri nets are a graphical formalism that is not very distant from the graphical representation of UML state machines.
Second, the UML formal specification mentions Petri nets as an analogy to both UML activity diagrams~\cite[p.\,285]{UML251} and state machine diagrams~\cite[p.\,313]{UML251}.

Explanations for this little support may be that Petri nets (and especially their extensions) are maybe less popular than, \eg{} \spin{}.
In addition, the lack of very well established model checking tools (in contrast to tools such as \uppaal{} or \spin{}) can be another explanation (\CPNTools{}, although well-known for CPNs, has a far from intuitive user interface).
Finally, Petri nets are a typically concurrent formalism, whereas UML state machines require a more global understanding: a local UML transition cannot be fired if some other transition in another region shall fire first (\eg{} because it has a higher priority).
This situation may be encoded into Petri nets, but usually using a rather cumbersome manner.

\subsection{Translation into the input language of \smv{} and \nusmv{}}\label{ss:smv}

A first translation into \smv{} is given by Kwon~\cite{Kwon00}: a formal semantics for UML statecharts is defined in the form of rule-rewriting systems,
and a translation approach is provided from the formalized semantics to the \smv{} model checker.
No detailed implementation is discussed.
This work fits in fact both into the translation approach and the operational semantics approach.

In~\cref{ss:asm}, we reviewed the work by Compton \etal{} in~\cite{CGHS00}, that translates UML state machines into ASMs.
In the same work, the authors actually then use \smv{} as the back-end model checker to automatically verify UML state machines.
The work first translates UML state machines into ASMs, which has been discussed before in \cref{ss:asm}.
Then, the \smv{} model checker is invoked to verify the \smv{} specification of a UML state machine.

Lam and Padget propose in~\cite{LP04} a symbolic encoding of UML statecharts, and invoke \nusmv{}~\cite{NuSMV} to perform the model 
checking.
Their work adopts a three-step procedure and uses $\phi$-calculus as an intermediate format for the translation.
They have implemented the translator from UML statecharts to $\phi$-calculus, and claim that the  implementation of a translator from $\phi$-calculus to the input language of \nusmv{} (named SC2PiCal) was ongoing---although we did not find any later updates on this.%

Beato \etal{} also provide in~\cite{BBCD05} a translation from UML diagrams to the input language of the \smv{} model checker.
Instead of focusing on just UML state machines, this work focuses on the collaborations of  different UML diagrams such as class diagrams, state machine diagrams and activity diagrams.
Noticing that high-level model designers may be unfamiliar with property languages used by model checkers (such as LTL and CTL), the authors also provide some aid in the form of a versatile assistant to guide users with their property writing.
This paper does not describe the detailed translation rules, but illustrates their translation procedure with an ATM machine example.
Hence, only the features appearing in that example are shown as ``\yes{}'' in \cref{table:translation:features}.

In~\cite{DJ08}, Dubrovin and Junttila first provide a compact symbolic encoding for UML state machines.
Its symbolic nature makes it suitable for an application to symbolic model checking, such as BDD-based (binary decision diagrams) or SAT-based bounded model checking.
The approach discusses event dispatching mechanisms, multi-object (asynchronous) communication as well as choice pseudostates---which are often left out in other approaches.
But on the other hand, some commonly considered constructs, such as history pseudostates, are not included in their formalization.
They perform a translation from the defined semantics to the input language of the \nusmv{}~\cite{NuSMV} symbolic model checker.
The detailed translation steps are not discussed in the paper, but an implementation SMUML~\cite{SMUML} has been provided, and some experiment results are reported in~\cite{DJ08}.
The symbolic step encoding itself is discussed in~\cite{DJH08}.

\subsection{Translation into process algebras}\label{ss:algebras}

Ng and Butler~\cite{NB02,NB03} propose to translate UML state machines into CSP and utilize the \fdr{} model  checker to proceed with the model checking procedure.
The priority mechanism is not encoded.
Thanks to the capabilities of \fdr{}, they are able to perform not only model checking but also trace refinement.
An extension of~\cite{NB03} is then considered in~\cite{YLWD05}.

Zhang and Liu~\cite{ZL10} provide an approach which translates UML state machines into 
\cspsharp{}, an extension of the CSP language, which serves as the input modeling language of \pat{}~\cite{SLDP09}.
Many aspects are considered; however, it is unclear whether the run-to-completion step is correctly encoded: the ``atomic'' construct used in the translation could ensure its correctness, but it does not seem to be used to encode a complete transition.
An implementation of the translator was done, and experimental results of the verification of UML state machines using \pat{} are presented.

Hansen \etal{}~\cite{HKLMP10} translate a subset of executable UML (xUML)~\cite{MB02} into the process algebraic specification language mCRL2~\cite{mCRL2,GMRUW06}.
Interestingly, they compare several definitions of the run-to-completion step, and show that this has an impact on the verification of properties.
In addition, the class diagram is translated together with the state machines.
Symbolic verification is performed using LTSmin~\cite{KLMPBD15}.
The xUML state machines subset considered in~\cite{HKLMP10} is close to that of UML~2.2.
The supported syntax includes concurrency and hierarchy, and event dispatching; history and final pseudostates are explicitly not supported, and fork/join do not seem to be either.

Then, Djaaboub \etal{} translate in~\cite{DKC15} UML state machines into flat state machines, and then ultimately into LOTOS (Language Of Temporal Ordering Specification)~\cite{BB87}.
A graph grammar is proposed for the translation, and their approach uses the meta-modeling tool AToM.
Few features only are considered, including composite states, and entry and exit behaviors.

Finally, Jacobs and Simpson translate in~\cite{JS15} a part of the activity diagram and state machine diagram syntax of the Systems Modeling Language (SysML).
SysML is close to UML (in fact the syntax of SysML state machines can be seen as a subset of UML behavioral state machines) and this work is therefore worth mentioning.
The UML state machine syntax supported by~\cite{JS15} is very limited, but its main interest is that the translation is performed in conjunction with activity diagrams.
The event queue seems well encoded.
As it is not exactly addressing UML state machines, we do not add this work to \cref{table:translation:features}.

\paragraph{Discussion}
According to~\cref{table:translation:features}, approaches that translate UML state machines to CSP do not seem more inclusive than other approaches in terms of supported syntax, with the exception of~\cite{ZL10}, that is the third most inclusive approach in the surveyed translation approaches.
However, it is unclear whether the run-to-completion step is properly encoded.
The proper encoding of the run-to-completion step requires some global view on the system, whereas CSP reasons in terms of parallel processes, which may be inherently incompatible with (or at least be quite inappropriate to model) the atomic, sequential nature of the run-to-completion step.
Although the proper encoding of the run-to-completion step using CSP is certainly possible, we believe that (similarly to Petri nets) it must be cumbersome, which explains its little support by the CSP-based translation approaches.

\subsection{Translation to PVS, KIV, B and Z}\label{ss:PVSBZ}

\paragraph{PVS}
Traoré~\cite{Traore00} and Aredo~\cite{Aredo00} proposed the same year in two independent works (though relatively close in spirit) to translate UML state machines into PVS (Prototype Verification System)~\cite{ORS92}.
Many non-trivial aspects of the syntax of UML are considered in these two works, such as shallow history pseudostates, submachines states and even time (in~\cite{Traore00}).

\paragraph{KIV}
In~\cite{BBKRT04}, Balser \etal{} use interactive verification in order to formally prove properties of UML state machines.
They use KIV~\cite{BRSS98}.
The property language is a variant of ITL (interval temporal logic)~\cite{Moszkowski85}.
The set of considered features is quite large, with composite states, internal transitions, inter-level transitions, forks and joins, variables, and a proper encoding of the run-to-completion step.
A quite complex example (\cite[Fig.1]{BBKRT04}) is given.
However, quite surprisingly, history pseudostates (both deep and shallow) are left out, and so are choice pseudostates, call and deferred events, as well as internal transitions.

\paragraph{B and extensions}
In~\cite{LS02}, Ledang and Souquières translate UML state machine diagrams into B specifications~\cite{Abrial96}.
The set of syntactic features is quite restricted, and no automation seems to be provided.
A simple lift control system is used for exemplification.

In~\cite{LCA04}, a subset of the UML-RSDS specifications~\cite{LFA02} is translated into~B.
The article mostly considers class diagrams, and state machines are only mentioned without being formally considered in the translation.
The focus is not mainly on state machines, and therefore this work is not integrated in \cref{table:translation:features}.
Note that UML-RSDS is now called AgileUML~\cite{LJT21}.\footnote{%
	\url{https://projects.eclipse.org/projects/modeling.agileuml}
}

Snook \etal{}~\cite{SSB10} provide an approach to translate UML models (including class diagrams and state machine diagrams) into UML-B diagrams~\cite{SB06,SB08}, which incorporate Event-B method into UML diagrams.
The idea of integrating UML with Event-B was introduced in~\cite{SB06} and improved by introducing an action modeling language $\mu$-B.
A translator U2B was developed, and later made an Eclipse plugin~\cite{SB08}, which is also part of the Rodin platform~\cite{ABHHMV10}.
Unfortunately, the translation in~\cite{SSB10} is not detailed enough, and therefore not integrated into \cref{table:translation:features}.

Refinement of systems through a combination of UML-B state machines and Event-B is also discussed in~\cite{BCESEGM13}.

\paragraph{Z}
In~\cite{KC02}, Kim and Carrington propose a model transformation based approach to transform UML state machine diagrams into Z specifications, but also a transformation from the latter formalism into the former.
The authors formalize a UML metamodel using Object-Z, and then propose the opposite formalization.
Only basic syntactic aspects are considered in the translation.

In~\cite{ZM04}, Zhan and Miao propose a formalization for UML state machines using~Z (extending notably \cite{MLL02}).
The model can then be translated to ``FREE'' (flattened regular expressions) models.
The run-to-completion step is carefully encoded, but many syntactic aspects of the UML are missing.
While the ultimate goal is testing, no automated translation seems to be available.
Testing UML state machines is further considered in~\cite{Zhan07}.

In~\cite{EE15}, El~Miloudi and Ettouhami propose an approach to translate UML state machine diagrams (in addition to class and sequence diagrams) into the Z notation~\cite{Spivey92}.
An originality is multi-view modeling, with a consistency check with class and sequence diagrams.
Few syntactic elements are considered in~\cite{EE15}, but signal events are considered.
Overall, the translation mechanism is described in a rather shallow manner, and no automatic translation engine seems to be available.

\paragraph{Discussion}
According to~\cref{table:translation:features}, the two works translating state machines to PVS~\cite{Traore00,Aredo00} are among the most complete when it comes to translating UML state machines into a target formalism (and \cite{Traore00} is actually the most complete work surveyed in the translation-based approaches, together with~\cite{JEJ04}).
They consider syntactic aspects that few other works support (submachine states, junctions, history pseudostates, run-to-completion step, and even time).
This is surprising as the UML version considered in their work is outdated (1.3) but also because these two works are among the two earliest.

\subsection{Formalization using institutions}\label{ss:institutions}

In~\cite{KMRG15}, Knapp \etal{} formalize ``simple'' (non-hierarchical) UML state machines using institutions.
In~\cite{KM17}, Knapp and Mossakowski do not only extend the former work to hierarchy, but also add sequence diagrams and composite structure diagrams.
The set of syntactic elements is rather limited, and no automated translation seems to be provided; however, a simple check using the Distributed Ontology, Model and Specification Language (DOL) is performed.

In~\cite{RBKR20}, UML state machine are first embedded into a logical framework (called ``$\mathcal{M}^\downarrow_{\mathcal{D}}$''), which is then mapped to \textsc{Casl} (Common Algebraic Specification Language)~\cite{Mosses04}.
As in~\cite{KMRG15}, only ``simple'' (flat) state machines are considered; the set of covered syntactic elements is not precisely given, but seems rather limited.
Notably, communication between state machines (and the associated event pool) is not considered.
However, an implementation of the translation is available, and the verification of a safety property on a simple counterexample is performed using the automated theorem prover \textsc{Spass}~\cite{WDFKSW09}.

\subsection{Summary}\label{ss:translation:summary}
\subsubsection{Summary of features}

We summarize the surveyed translation-based approaches in \cref{table:translation:features}, ordered by target formalism.
We give from left to right
	the number of citations (see below),
	the UML version,
	the target formalism
	and then we record the features considered by each work, \viz{}
	the orthogonal and submachine states,
	the fork/join, junction, choice, shallow history, deep history pseudostates,
	the entry/exit points,
	the entry/exit behaviors,
	the internal, inter-level and completion transitions,
	the run-to-completion step,
	the use of variables,
	the proper handling of deferred events,
	the handling of time,
	and
	the ability to have communicating state machines, \ie{} several state machines communicating with each other, \eg{} using synchronization.
For space consideration, we remove the features that are commonly supported by all approaches, typically simple states, external transitions, initial pseudostate and final states.

\paragraph{Evaluating the syntax support}
The symbol ``$\yes$'' denotes the fact that a syntactic feature is supported, ``$\no$'' means the feature is not supported, ``$\partially$'' means the featured is discussed in the paper, but is not thoroughly solved.
For example, for ``conflict/priority'', some works considered conflict among enabled transitions, but did not discuss conflict due to deferred events.
In this case, we regard the features to be partially supported.
Some information could not be retrieved from some papers, in which case we mark it by ``$\unclear$''.
At the bottom of the table, we count the approaches handling each feature (we count 1 when ``$\yes$'', 0 when ``$\no$'', 0.5 when ``$\partially$'' and 0.25 when ``$\unclear$''); we hence derive the percentage of works handling that feature.
The right-most column (``sum'') of \cref{table:translation:features} also counts for each approach the number of supported features, by counting the supported features with the aforementioned conventions (1 when ``$\yes$'', 0 when ``$\no$'', 0.5 when ``$\partially$'' and 0.25 when ``$\unclear$'').
This is certainly not an absolute way to compare approaches with each other (see threats to validity below), but rather a way to quickly identify more complete approaches as opposed to approaches supporting very few syntactic features.\label{newtext:scores}

We can conclude from \cref{table:translation:features} that, in the translation-based approaches, the least supported features are entry/exit points, submachine states, time, and deferred events.
Internal transitions are discarded from most works formalizing UML, which may come as a surprise as they do not seem to pose particular theoretical problems.
In contrast, the most supported features are orthogonal states, entry/exit behaviors, and completion and inter-level transitions.
But quite disappointingly, with the exception of orthogonal states (68\,\%) and entry/exit behaviors (52\,\%), none of the UML state machines elements are supported by more than half of the surveyed works.

\paragraph{Evaluating the popularity}
Finally, we also tabulate the number of citations of each work (given by Google Scholar, as of early April~2022) in \cref{table:translation:features}.
While this measure is not perfect, and while earlier works have obviously more been cited since their publication than more recent works, it gives an (approximate) measure of the popularity of each work.
Only four of the surveyed works reached over 250 citations (with a maximum of~444): \cite{LMM99,LMM99Spin,SKM01,KMR02}, all prior to~2002, and from only two groups of authors: Latella, Majzik and Massink on the one hand \cite{LMM99,LMM99Spin}, and Knapp and~Merz on the other hand~\cite{SKM01,KMR02}.
They also consider rather restricted subsets of the UML syntax.
An explanation for this popularity may however come from the target language: 
three of them target \promela{}, the input language of \spin{},
while the last one~\cite{KMR02} is associated to a tool (\hugort{}) the development of which is still active nowadays (see \cref{tab:tools:availability}).\label{newtext:citations:trans}

\paragraph{Threats to validity}\label{newtext:threats-scores}
We briefly discuss possible threats to validity regarding \cref{table:translation:features}.
A first potential issue is that knowing exactly the set of features supported by each work is not always easy.
Most works did not provide an exact summary, leading us to have a detailed look at sometimes subtle semantic definitions, or even at the examples more or less formally detailed in each paper, in order to gather the set of supported syntactic features.
This also explains a number of ``$\partially$'' and ``$\unclear$'' cells in \cref{table:translation:features}.

A second potential issue is the ``sum'' score: while we believe it is useful in order to quickly identify rather complete approaches, it should not be taken too literally:
for example, from \cref{table:translation:features}, \cite{DMY02} has sum~10 while \cite{KMR02} has sum~9, but we do not necessarily mean that \cite{DMY02} is a ``better'' approach than~\cite{KMR02}.
	It solely means that \cite{DMY02} supports one more feature: both works \cite{DMY02,KMR02} are certainly more ``complementary'' to each other, rather than one strictly better than the other.
In addition, note that we counted exactly ``1'' for each feature to compute the sum; this is debatable, as some features may be considered as more important than others.
(And the relative importance of the syntactic features of UML state machines certainly goes beyond the scope of this survey.)
Finally, counting 0.5 for ``$\partially$'' and 0.25 for ``$\unclear$'' can again be seen as a somehow arbitrary choice.
Note however that a different method (\eg{} 0.25 for ``$\partially$'' and 0 for ``$\unclear$'') would not change significantly the order of the respective sums.

\subsubsection{Tool support}

We review the tool support offered by the translation-based approaches in \cref{table:translation:tools}.
We give from left to right
	the UML version,
	the target formalism,
	the model checking tool (if any),
	the kind of verification offered,
	and the tool responsible to perform the translation (if any).
From \cref{table:translation:tools}, we see that, while many approaches offer a tool support, some do not---which can be seen as a somehow debatable choice considering the highly applied motivation of this field of research.
In addition, a few approaches aim at translating UML into some existing model checkers, but do not provide any automated software to do so.
This is again a very debatable choice.

We will review all tools from a user point of view in \cref{section:tools}.

\subsubsection{Discussion}\label{sss:discussion:features}

Clearly, a first disappointing conclusion is that most works consider a quite restricted subset of the syntax of state machines.
No work considers more than 11/19 features which gives 58\,\%, hence not much more than one half.
Conversely, with the exception of orthogonal composite states (68\,\%) and entry/exit behaviors (52\,\%), no feature is supported by more than 50\,\% of the approaches.
The strong support of a particular feature does not seem to be specifically correlated with neither the fact that it is older (introduced in earlier versions of the UML), nor the fact that it can be substituted easily by some other features.

The least supported feature is the entry/exit points (2\,\%, a single approach), which is highly surprising as these entry/exit points do not bring any difficulty to translate to other languages in a relatively easy fashion; in addition, they can turn really useful from a modeling perspective.
More worrying is the fact that the run-to-completion step (a central notion in UML) is not correctly supported by most approaches (only 29\,\%).

Another surprising point is that more recent approaches are not necessarily the best; in fact, the most complete approaches (\ie{} \cite{Traore00,JEJ04}) are among the oldest.

Finally, two recent works~\cite{MCSK17,MCSK19} are rather shallow in terms of the syntax considered, but are a verified translation, which brings some additional guarantees into the confidence one can have in the result.

\subsubsection{Common advantages and drawbacks of translation-based approaches}

The translation approaches aim at utilizing the automatic verification ability of different model 
checkers.
However, translation-based approaches may suffer from the following defects:
\begin{enumerate}
	\item Due to the semantic gaps, it may be hard to translate some syntactic features of UML state machines, introducing sometimes additional but undesired behaviors.
	For example in~\cite{ZL10}, extra events have to be added to each process so as to model exit behaviors of orthogonal composite states. %
	
	\item For the verification, translation approaches heavily depend on the tool support of the target formal languages.
	Furthermore, the additional behaviors introduced during the translation may significantly slow down the verification.
	These additional behaviors may come in the form of additional steps required by the translation; for example, in automata-based formalisms, some (internal) automata transitions may be added, which do not modify the overall UML state machine execution, but still add some extra transitions to be explored by the model checker; this may blow up in case of different automata in parallel.
	(Similar behaviors can happen in other formalisms such as Petri nets, process algebras, etc.)\label{newtext:extrabehaviors}
	In addition, optimizations and reduction techniques (such as partial order reduction) may not apply in order to preserve the semantics of the original model.
	
	\item Lastly, when a counterexample is found by the verification tool, it is hard to map it to the original state machine execution, especially when state space reduction techniques are used.
		One of the exceptions is \cite{CGHS00,SCH01}, where a counterexample can be exhibited in a visual fashion.
\end{enumerate}

Note that a direct implementation based on an operational semantics may solve some of the aforementioned issues.

\ifdefined\VersionLong
	\begin{landscape}

	\thispagestyle{empty}
\fi
\begin{table}
	\caption{UML state machine features supported by each translation approach}
	\setlength{\tabcolsep}{2pt}
	\scalebox{\ifdefined\VersionLong .6\else .48\fi}{

\begin{tabular}{ |c|c|c|c|c|c|c|c|c|c|c|c|c|c|c|c|c|c|c|c|c|c|}
	\hline
	\rowcolor{colorHead}
	\bfseries{}Approach & nb & UML & Target & 			\multicolumn{2}{c|}{\cellcolor{colorHead}States} & 			\multicolumn{6}{c|}{\cellcolor{colorHead}Pseudostates} & 			Entry/exit & 			\multicolumn{3}{c|}{\cellcolor{colorHead}Transitions} & 			RTC & 			Variables & 			Deferred & 			Time & 			Multiple & 			Sum 			
	\\
	\rowcolor{colorHead}& cit. & v & & 			ortho & subm & 			fk/jn & junct. & choice & sH & dH & en/ex & 			behv & 			intern & interl & compl & 			& 			& 			events & 			& 			charts 			& 		\\
	\hline
	Börger \etal{} (2000) \cite{BCR00} & 	\cellcolor{blue!32}146 & 1.3 & ASM & \partially{} & \partially{} & \partially{} & \partially{} & \no{} & \yes{} & \yes{} & \no{} & \yes{} & \yes{} & \partially{} & \yes{} & \yes{} & \no{} & \yes{} & \no{} & \no{} &\cellcolor{blue!55}9.5
	\\
	\hline
	Compton \etal{} (2000) \cite{CGHS00} & 	\cellcolor{blue!12}55 & 1.3 & ASM & \yes{} & \no{} & \no{} & \no{} & \no{} & \no{} & \no{} & \no{} & \yes{} & \no{} & \yes{} & \yes{} & \no{} & \no{} & \unclear{} & \no{} & \no{} &\cellcolor{blue!25}4.25
	\\
	\hline
	Jürjens (2002) \cite{Jurjens02} & 	\cellcolor{blue!12}56 & 1.4 & ASM & \yes{} & \no{} & \no{} & \no{} & \no{} & \no{} & \no{} & \no{} & \yes{} & \yes{} & \no{} & \partially{} & \no{} & \no{} & \no{} & \no{} & \yes{} &\cellcolor{blue!26}4.5
	\\
	\hline
	Börger \etal{} (2003) \cite{BRC03} & 	\cellcolor{blue!3}16 & 1.3 & ASM & \yes{} & \no{} & \no{} & \no{} & \no{} & \yes{} & \yes{} & \no{} & \yes{} & \unclear{} & \unclear{} & \yes{} & \unclear{} & \unclear{} & \yes{} & \no{} & \no{} &\cellcolor{blue!35}6.0
	\\
	\hline
	Jin \etal{} (2004) \cite{JEJ04} & 	\cellcolor{blue!7}34 & 1.5 & OMA & \yes{} & \no{} & \yes{} & \yes{} & \no{} & \yes{} & \yes{} & \no{} & \yes{} & \yes{} & \yes{} & \yes{} & \yes{} & \no{} & \yes{} & \no{} & \no{} &\cellcolor{blue!64}11.0
	\\

	\hline
	\hline
	Kuske (2001) \cite{Kuske01} & 	\cellcolor{blue!29}130 & 1.3 & graph transformation & \yes{} & \no{} & \unclear{} & \no{} & \no{} & \no{} & \no{} & \no{} & \indirectly{} & \indirectly{} & \no{} & \no{} & \yes{} & \no{} & \unclear{} & \no{} & \no{} &\cellcolor{blue!26}4.5
	\\
	\hline
	Kong \etal{} (2009) \cite{KZDX09} & 	\cellcolor{blue!11}51 & 2.0 & graph transformation & \yes{} & \no{} & \yes{} & \yes{} & \yes{} & \yes{} & \yes{} & \no{} & \no{} & \no{} & \yes{} & \yes{} & \unclear{} & \yes{} & \no{} & \no{} & \yes{} &\cellcolor{blue!60}10.25
	\\

	\hline
	\hline
	Latella \etal{} (1999) \cite{LMM99} & 	\cellcolor{blue!71}317 & $<$1.3 & EHA & \yes{} & \no{} & \yes{} & \no{} & \no{} & \no{} & \no{} & \no{} & \no{} & \no{} & \yes{} & \no{} & \no{} & \no{} & \no{} & \no{} & \no{} &\cellcolor{blue!17}3.0
	\\
	\hline
	Gnesi \etal{} (1999) \cite{GLM99} & 	\cellcolor{blue!25}112 & $<$1.3 & EHA/LTS & \yes{} & \no{} & \yes{} & \no{} & \no{} & \no{} & \no{} & \no{} & \no{} & \no{} & \yes{} & \no{} & \no{} & \no{} & \no{} & \no{} & \no{} &\cellcolor{blue!17}3.0
	\\
	\hline
	Latella \etal{} (1999) \cite{LMM99Spin} & 	\cellcolor{blue!100}444 & 1.1 & EHA/\promela{} & \yes{} & \no{} & \yes{} & \no{} & \no{} & \no{} & \no{} & \no{} & \no{} & \no{} & \yes{} & \no{} & \no{} & \no{} & \no{} & \no{} & \no{} &\cellcolor{blue!17}3.0
	\\

	\hline
	\hline
	Schäfer \etal{} (2001) \cite{SKM01} & 	\cellcolor{blue!70}313 & 1.4 & \promela{} & \yes{} & \no{} & \yes{} & \no{} & \yes{} & \no{} & \no{} & \no{} & \yes{} & \no{} & \yes{} & \yes{} & \yes{} & \no{} & \no{} & \no{} & \no{} &\cellcolor{blue!41}7.0
	\\
	\hline
	Jussila \etal{} (2006) \cite{JDJLP06} & 	\cellcolor{blue!20}89 & 1.4 & \promela & \yes{} & \no{} & \no{} & \no{} & \yes{} & \no{} & \no{} & \no{} & \no{} & \no{} & \no{} & \yes{} & \no{} & \yes{} & \yes{} & \no{} & \yes{} &\cellcolor{blue!35}6.0
	\\
	\hline
	Carlsson \etal{} (2009) \cite{CJ09} & 	\cellcolor{blue!1}7 & 2.0 (?) & \promela{} & \no{} & \no{} & \no{} & \no{} & \no{} & \no{} & \no{} & \no{} & \no{} & \no{} & \no{} & \no{} & \no{} & \no{} & \no{} & \yes{} & \yes{} &\cellcolor{blue!11}2.0
	\\

	\hline
	\hline
	Knapp \etal{} (2002) \cite{KMR02} & 	\cellcolor{blue!58}261 & 1.4 & TA & \yes{} & \no{} & \yes{} & \yes{} & \no{} & \no{} & \no{} & \no{} & \yes{} & \no{} & \no{} & \yes{} & \yes{} & \yes{} & \no{} & \yes{} & \yes{} &\cellcolor{blue!52}9.0
	\\
	\hline
	David \etal{} (2002) \cite{DMY02} & 	\cellcolor{blue!36}163 & 1.4 (?) & TA & \yes{} & \yes{} & \yes{} & \yes{} & \no{} & \yes{} & \no{} & \no{} & \no{} & \no{} & \yes{} & \yes{} & \no{} & \yes{} & \no{} & \yes{} & \yes{} &\cellcolor{blue!58}10.0
	\\
	\hline
	Ober \etal{} (2006) \cite{OGO06} & 	\cellcolor{blue!31}138 & 2.0 & communicating ext.\ TA & \unclear{} & \no{} & \no{} & \no{} & \yes{} & \no{} & \no{} & \no{} & \unclear{} & \no{} & \no{} & \no{} & \yes{} & \yes{} & \no{} & \yes{} & \unclear{} &\cellcolor{blue!27}4.75
	\\

	\hline
	\hline
	Pettit IV \etal{} (2000) \cite{PG00} & 	\cellcolor{blue!14}63 & $\leq$1.3 & CPN & \no{} & \no{} & \no{} & \no{} & \no{} & \no{} & \no{} & \no{} & \no{} & \no{} & \no{} & \no{} & \no{} & \no{} & \no{} & \no{} & \yes{} &\cellcolor{blue!5}1.0
	\\
	\hline
	Baresi \etal{} (2001) \cite{BP01} & 	\cellcolor{blue!17}78 & $\leq$1.3 & HLPN & \no{} & \no{} & \no{} & \no{} & \no{} & \no{} & \no{} & \no{} & \no{} & \no{} & \no{} & \no{} & \no{} & \no{} & \no{} & \no{} & \no{} &\cellcolor{blue!0}0.0
	\\
	\hline
	Saldhana \etal{} (2001) \cite{SSH01} & 	\cellcolor{blue!18}84 & 1.2? & objet Petri nets & \yes{} & \no{} & \yes{} & \no{} & \no{} & \no{} & \no{} & \no{} & \yes{} & \yes{} & \no{} & \no{} & \no{} & \no{} & \yes{} & \no{} & \yes{} &\cellcolor{blue!35}6.0
	\\
	\hline
	Merseguer \etal{} (2002) \cite{MCBD02} & 	\cellcolor{blue!26}116 & 1.4 & GSPNs & \no{} & \no{} & \no{} & \no{} & \no{} & \no{} & \no{} & \no{} & \yes{} & \yes{} & \no{} & \yes{} & \unclear{} & \no{} & \yes{} & \no{} & \no{} &\cellcolor{blue!25}4.25
	\\
	\hline
	Trowitzsch \etal{} (2005) \cite{TZ05} & 	\cellcolor{blue!4}22 & 2.0 & SPN & \partially{} & \no{} & \yes{} & \yes{} & \yes{} & \no{} & \no{} & \no{} & \yes{} & \no{} & \no{} & \no{} & \no{} & \yes{} & \no{} & \yes{} & \no{} &\cellcolor{blue!38}6.5
	\\
	\hline
	Choppy \etal{} (2011) \cite{CKZ11} & 	\cellcolor{blue!13}59 & 2.2 (?) & HCPN & \partially{} & \no{} & \no{} & \yes{} & \yes{} & \no{} & \no{} & \no{} & \yes{} & \no{} & \no{} & \no{} & \no{} & \no{} & \no{} & \no{} & \yes{} &\cellcolor{blue!26}4.5
	\\
	\hline
	André \etal{} (2012) \cite{ACK12} & 	\cellcolor{blue!5}23 & 2.2 & CPN & \no{} & \no{} & \no{} & \no{} & \no{} & \yes{} & \partially{} & \no{} & \yes{} & \yes{} & \yes{} & \yes{} & \no{} & \yes{} & \no{} & \no{} & \no{} &\cellcolor{blue!38}6.5
	\\
	\hline
	André \etal{} (2016) \cite{ABC16} & 	\cellcolor{blue!2}12 & 2.5 beta 1 & CPN & \yes{} & \partially{} & \yes{} & \no{} & \no{} & \yes{} & \partially{} & \no{} & \yes{} & \yes{} & \yes{} & \yes{} & \yes{} & \yes{} & \no{} & \no{} & \no{} &\cellcolor{blue!58}10.0
	\\
	\hline
	Kumar \etal{} (2017) \cite{KST17} & 	\cellcolor{blue!3}15 & $\leq$ 2.5 & Petri nets & \yes{} & \no{} & \yes{} & \no{} & \no{} & \no{} & \no{} & \no{} & \unclear{} & \no{} & \no{} & \no{} & \no{} & \no{} & \no{} & \no{} & \no{} &\cellcolor{blue!13}2.25
	\\
	\hline
	Meghzili \etal{} (2017) \cite{MCSK17} & 	\cellcolor{blue!3}16 & $\leq$ 2.5 & CPN & \yes{} & \no{} & \no{} & \no{} & \no{} & \no{} & \no{} & \no{} & \yes{} & \no{} & \no{} & \no{} & \no{} & \unclear{} & \no{} & \no{} & \no{} &\cellcolor{blue!13}2.25
	\\
	\hline
	Lyazidi \etal{} (2019) \cite{LM19} & 	\cellcolor{blue!0}3 & $\leq$ 2.5.1 & Petri nets & \no{} & \no{} & \yes{} & \yes{} & \yes{} & \no{} & \no{} & \no{} & \no{} & \no{} & \no{} & \no{} & \no{} & \no{} & \no{} & \unclear{} & \no{} &\cellcolor{blue!19}3.25
	\\

	\hline
	\hline
	Kwon (2000) \cite{Kwon00} & 	\cellcolor{blue!18}81 & 1.3 & \smv{} & \yes{} & \no{} & \no{} & \no{} & \no{} & \yes{} & \no{} & \no{} & \no{} & \no{} & \yes{} & \no{} & \no{} & \no{} & \no{} & \no{} & \no{} &\cellcolor{blue!17}3.0
	\\
	\hline
	Lam \etal{} (2004) \cite{LP04} & 	\cellcolor{blue!9}42 & $\leq$1.5 & \nusmv{} & \no{} & \no{} & \no{} & \no{} & \no{} & \no{} & \no{} & \no{} & \no{} & \unclear{} & \unclear{} & \no{} & \no{} & \no{} & \no{} & \no{} & \yes{} &\cellcolor{blue!5}1.0
	\\
	\hline
	Beato \etal{} (2005) \cite{BBCD05} & 	\cellcolor{blue!15}68 & $<$1.3 & \smv{} & \yes{} & \no{} & \no{} & \no{} & \yes{} & \no{} & \no{} & \no{} & \yes{} & \no{} & \yes{} & \no{} & \no{} & \yes{} & \no{} & \no{} & \unclear{} &\cellcolor{blue!30}5.25
	\\
	\hline
	Dubrovin \etal{} (2008) \cite{DJ08} & 	\cellcolor{blue!14}64 & $\leq$2.1.1 & \nusmv{} & \yes{} & \no{} & \no{} & \no{} & \yes{} & \no{} & \no{} & \no{} & \no{} & \no{} & \yes{} & \yes{} & \yes{} & \yes{} & \yes{} & \no{} & \yes{} &\cellcolor{blue!47}8.0
	\\

	\hline
	\hline
	Ng \etal{} (2002) \cite{NB02} & 	\cellcolor{blue!8}37 & 1.3 & CSP & \no{} & \no{} & \no{} & \no{} & \yes{} & \no{} & \no{} & \no{} & \no{} & \no{} & \no{} & \no{} & \no{} & \yes{} & \no{} & \no{} & \yes{} &\cellcolor{blue!17}3.0
	\\
	\hline
	Ng \etal{} (2003) \cite{NB03} & 	\cellcolor{blue!18}82 & 1.4 & CSP & \yes{} & \no{} & \no{} & \no{} & \yes{} & \no{} & \no{} & \no{} & \yes{} & \no{} & \partially{} & \yes{} & \no{} & \yes{} & \no{} & \no{} & \yes{} &\cellcolor{blue!38}6.5
	\\
	\hline
	Zhang and Liu (2010) \cite{ZL10} & 	\cellcolor{blue!14}63 & 2.2 & \cspsharp{} & \yes{} & \yes{} & \yes{} & \no{} & \yes{} & \yes{} & \no{} & \yes{} & \yes{} & \no{} & \yes{} & \yes{} & \unclear{} & \yes{} & \no{} & \no{} & \no{} &\cellcolor{blue!60}10.25
	\\
	\hline
	Hansen \etal{} (2010) \cite{HKLMP10} & 	\cellcolor{blue!13}59 & 2.2 & mCRL2 & \yes{} & \no{} & \no{} & \no{} & \no{} & \no{} & \no{} & \no{} & \no{} & \no{} & \no{} & \yes{} & \yes{} & \no{} & \no{} & \no{} & \no{} &\cellcolor{blue!17}3.0
	\\
	\hline
	Djaaboub \etal{} (2015) \cite{DKC15} & 	\cellcolor{blue!0}2 & 2.0 & LOTOS & \yes{} & \no{} & \no{} & \no{} & \no{} & \no{} & \no{} & \no{} & \yes{} & \no{} & \yes{} & \no{} & \no{} & \no{} & \no{} & \no{} & \no{} &\cellcolor{blue!17}3.0
	\\

	\hline
	\hline
	Aredo (2000) \cite{Aredo00} & 	\cellcolor{blue!5}24 & 1.3 & PVS & \yes{} & \yes{} & \yes{} & \yes{} & \yes{} & \yes{} & \yes{} & \no{} & \yes{} & \no{} & \no{} & \yes{} & \no{} & \yes{} & \no{} & \no{} & \no{} &\cellcolor{blue!58}10.0
	\\
	\hline
	Traoré (2000) \cite{Traore00} & 	\cellcolor{blue!10}47 & 1.3 & PVS & \yes{} & \no{} & \yes{} & \yes{} & \yes{} & \yes{} & \yes{} & \no{} & \no{} & \no{} & \yes{} & \yes{} & \yes{} & \yes{} & \no{} & \yes{} & \no{} &\cellcolor{blue!64}11.0
	\\
	\hline
	Kim \etal{} (2002) \cite{KC02} & 	\cellcolor{blue!5}24 & 1.3 & Z & \yes{} & \no{} & \no{} & \no{} & \no{} & \yes{} & \yes{} & \no{} & \yes{} & \yes{} & \no{} & \no{} & \no{} & \no{} & \yes{} & \no{} & \no{} &\cellcolor{blue!35}6.0
	\\
	\hline
	Ledang \etal{} (2002) \cite{LS02} & 	\cellcolor{blue!19}88 & 1.2? & B & \no{} & \no{} & \no{} & \no{} & \no{} & \no{} & \no{} & \no{} & \yes{} & \no{} & \yes{} & \no{} & \unclear{} & \no{} & \yes{} & \no{} & \yes{} &\cellcolor{blue!25}4.25
	\\
	\hline
	Balser \etal{} (2004) \cite{BBKRT04} & 	\cellcolor{blue!15}68 & 1.5 & KIV & \yes{} & \no{} & \yes{} & \yes{} & \no{} & \no{} & \no{} & \no{} & \yes{} & \no{} & \yes{} & \yes{} & \yes{} & \yes{} & \no{} & \no{} & \no{} &\cellcolor{blue!47}8.0
	\\
	\hline
	Zhan \etal{} (2004) \cite{ZM04} & 	\cellcolor{blue!3}14 & 2.0 & Z & \yes{} & \unclear{} & \no{} & \no{} & \no{} & \no{} & \no{} & \no{} & \no{} & \yes{} & \yes{} & \yes{} & \yes{} & \yes{} & \no{} & \no{} & \no{} &\cellcolor{blue!36}6.25
	\\
	\hline
	El Miloudi \etal{} (2015) \cite{EE15} & 	\cellcolor{blue!1}6 & 2.4.1 & Z & \no{} & \no{} & \unclear{} & \no{} & \no{} & \no{} & \no{} & \no{} & \unclear{} & \no{} & \no{} & \no{} & \no{} & \no{} & \no{} & \no{} & \no{} &\cellcolor{blue!2}0.5
	\\

	\hline
	\hline
	Knapp \etal{} (2017) \cite{KM17} & 	\cellcolor{blue!1}8 & 2.5 & interactions & \no{} & \unclear{} & \no{} & \yes{} & \no{} & \no{} & \no{} & \no{} & \no{} & \unclear{} & \no{} & \yes{} & \unclear{} & \yes{} & \no{} & \no{} & \no{} &\cellcolor{blue!22}3.75
	\\
	\hline
	Rosenberger \etal{} (2021) \cite{RBKR20} & 	\cellcolor{blue!0}1 & 2.5.1 & interactions & \no{} & \no{} & \no{} & \no{} & \no{} & \no{} & \no{} & \no{} & \no{} & \no{} & \no{} & \no{} & \no{} & \unclear{} & \no{} & \no{} & \no{} &\cellcolor{blue!1}0.25
	\\
	\hline
	\hline
	\cellcolor{colorHead}Features supported & -  & -  & -  & 30.75 & 4.5 & 18.0 & 11.5 & 14.0 & 12.0 & 8.0 & 1.0 & 23.75 & 10.25 & 20.0 & 21.5 & 13.25 & 18.5 & 9.5 & 6.25 & 14.5 & 17

	\\
	\hline
	\cellcolor{colorHead}\% & -  & -  & -  & \cellcolor{blue!68}68\,\% & \cellcolor{blue!10}10\,\% & \cellcolor{blue!40}40\,\% & \cellcolor{blue!25}25\,\% & \cellcolor{blue!31}31\,\% & \cellcolor{blue!26}26\,\% & \cellcolor{blue!17}17\,\% & \cellcolor{blue!2}2\,\% & \cellcolor{blue!52}52\,\% & \cellcolor{blue!22}22\,\% & \cellcolor{blue!44}44\,\% & \cellcolor{blue!47}47\,\% & \cellcolor{blue!29}29\,\% & \cellcolor{blue!41}41\,\% & \cellcolor{blue!21}21\,\% & \cellcolor{blue!13}13\,\% & \cellcolor{blue!32}32\,\% & 100\,\%

	\\

	\hline

\end{tabular}

}
	\label{table:translation:features}
\end{table}
\ifdefined\VersionLong
	\end{landscape}
\fi
\ifdefined\VersionLong
	\begin{landscape}

	\thispagestyle{empty}
\fi

\begin{table}
	\caption{Translation approach: Tool support}
	\setlength{\tabcolsep}{2pt}
	\scalebox{.6}{

\begin{tabular}{ |c|c|c|c|c|c|}
	\hline
	\rowcolor{colorHead}
	\bfseries{}Approach & UML v & Target & Verification tool & Verification & Translation tool
	\\
	\hline
	Börger \etal{} (2000) \cite{BCR00} & 1.3 & ASM & \no{} & \no{} & \no{}
	\\
	\hline
	Compton \etal{} (2000) \cite{CGHS00} & 1.3 & ASM & \cellColorYes{}\smv{} & \cellColorYes{}model checking & \cellColorYes{}\veriUML{}
	\\
	\hline
	Jürjens (2002) \cite{Jurjens02} & 1.4 & ASM & \no{} & \no{} & \no{}
	\\
	\hline
	Börger \etal{} (2003) \cite{BRC03} & 1.3 & ASM & \no{} & \no{} & \no{}
	\\
	\hline
	Jin \etal{} (2004) \cite{JEJ04} & 1.5 & OMA & \no{} & \no{} & \no{}
	\\

	\hline
	\hline
	Kuske (2001) \cite{Kuske01} & 1.3 & graph transformation & \no{} & \no{} & \no{}
	\\
	\hline
	Kong \etal{} (2009) \cite{KZDX09} & 2.0 & graph transformation & \no{} & \no{} & \no{}
	\\

	\hline
	\hline
	Latella \etal{} (1999) \cite{LMM99} & $<$1.3 & EHA & \no{} & \no{} & \no{}
	\\
	\hline
	Gnesi \etal{} (1999) \cite{GLM99} & $<$1.3 & EHA/LTS & \cellColorYes{}\jack{} & \cellColorYes{}model checking with ACTL & \no{}
	\\
	\hline
	Latella \etal{} (1999) \cite{LMM99Spin} & 1.1 & EHA/\promela{} & \cellColorYes{}\spin{} & \cellColorYes{}model checking with ACTL & \yes{}
	\\

	\hline
	\hline
	Schäfer \etal{} (2001) \cite{SKM01} & 1.4 & \promela{} & \cellColorYes{}\spin{} & \cellColorYes{}model checking & \cellColorYes{}\hugo{}
	\\
	\hline
	Jussila \etal{} (2006) \cite{JDJLP06} & 1.4 & \promela & \cellColorYes{}\spin{} & \cellColorYes{}model checking & \cellColorYes{}\proco{}
	\\
	\hline
	Carlsson \etal{} (2009) \cite{CJ09} & 2.0 (?) & \promela{} & \cellColorYes{}\spin{} & \cellColorYes{}model checking & \cellColorYes{}RSARTE
	\\

	\hline
	\hline
	Knapp \etal{} (2002) \cite{KMR02} & 1.4 & TA & \cellColorYes{}\uppaal{} & \cellColorYes{}subset of TCTL & \cellColorYes{}\hugort{}
	\\
	\hline
	David \etal{} (2002) \cite{DMY02} & 1.4 (?) & TA & \cellColorYes{}\uppaal{} & \cellColorYes{}subset of TCTL & \unclear{}
	\\
	\hline
	Ober \etal{} (2006) \cite{OGO06} & 2.0 & communicating ext.\ TA & \cellColorYes{}IF & \cellColorYes{}simulation and verification & \yes{}
	\\

	\hline
	\hline
	Pettit IV \etal{} (2000) \cite{PG00} & $\leq$1.3 & CPN & \cellColorYes{}DesignCPN & \cellColorYes{}deadlock / statistical analysis & \no{}
	\\
	\hline
	Baresi \etal{} (2001) \cite{BP01} & $\leq$1.3 & HLPN & \no{} & \no{} & \no{}
	\\
	\hline
	Saldhana \etal{} (2001) \cite{SSH01} & 1.2? & objet Petri nets & \no{} & \no{} & \no{}
	\\
	\hline
	Merseguer \etal{} (2002) \cite{MCBD02} & 1.4 & GSPNs & \no{} & \no{} & \no{}
	\\
	\hline
	Trowitzsch \etal{} (2005) \cite{TZ05} & 2.0 & SPN & \no{} & \no{} & \no{}
	\\
	\hline
	Choppy \etal{} (2011) \cite{CKZ11} & 2.2 (?) & HCPN & \cellColorYes{}\CPNAMI{} & \cellColorYes{}LTL/CTL & \yes{}
	\\
	\hline
	André \etal{} (2012) \cite{ACK12} & 2.2 & CPN & \cellColorYes{}\CPNTools{} & \cellColorYes{}model checking & \no{}
	\\
	\hline
	André \etal{} (2016) \cite{ABC16} & 2.5 beta 1 & CPN & \cellColorYes{}\CPNTools{} & \cellColorYes{}model checking & \cellColorYes{}prototype
	\\
	\hline
	Kumar \etal{} (2017) \cite{KST17} & $\leq$ 2.5 & Petri nets & \no{} & \cellColorYes{}probabilistic model checking & \no{}
	\\
	\hline
	Meghzili \etal{} (2017) \cite{MCSK17} & $\leq$ 2.5 & CPN & \cellColorYes{}none (\CPNTools{} envisioned) & \cellColorYes{}model checking & \cellColorYes{}Scala
	\\
	\hline
	Lyazidi \etal{} (2019) \cite{LM19} & $\leq$ 2.5.1 & Petri nets & \cellColorYes{}TiNA & \cellColorYes{}model checking & \cellColorYes{}prototype
	\\

	\hline
	\hline
	Kwon (2000) \cite{Kwon00} & 1.3 & \smv{} & \cellColorYes{}\smv{} & \cellColorYes{}model checking & \unclear{}
	\\
	\hline
	Lam \etal{} (2004) \cite{LP04} & $\leq$1.5 & \nusmv{} & \cellColorYes{}\nusmv{} & \cellColorYes{}model checking & \cellColorYes{}SC2PiCal (?)
	\\
	\hline
	Beato \etal{} (2005) \cite{BBCD05} & $<$1.3 & \smv{} & \cellColorYes{}\smv{} & \cellColorYes{}model checking & \cellColorYes{}\tabu{}
	\\
	\hline
	Dubrovin \etal{} (2008) \cite{DJ08} & $\leq$2.1.1 & \nusmv{} & \cellColorYes{}\nusmv{} & \cellColorYes{}model checking & \yes{}
	\\

	\hline
	\hline
	Ng \etal{} (2002) \cite{NB02} & 1.3 & CSP & \cellColorYes{}\fdr{} & \cellColorYes{}model checking / refinement & \yes{}
	\\
	\hline
	Ng \etal{} (2003) \cite{NB03} & 1.4 & CSP & \cellColorYes{}\fdr{} & \cellColorYes{}model checking / refinement & \yes{}
	\\
	\hline
	Zhang and Liu (2010) \cite{ZL10} & 2.2 & \cspsharp{} & \cellColorYes{}\pat{} & \cellColorYes{}model checking & \unclear{}
	\\
	\hline
	Hansen \etal{} (2010) \cite{HKLMP10} & 2.2 & mCRL2 & \cellColorYes{}LTSmin & \cellColorYes{}model checking & \no{}
	\\
	\hline
	Djaaboub \etal{} (2015) \cite{DKC15} & 2.0 & LOTOS & \no{} & \no{} & \cellColorYes{}graph grammar
	\\

	\hline
	\hline
	Aredo (2000) \cite{Aredo00} & 1.3 & PVS & \cellColorYes{}PVS & \cellColorYes{}model checking / theorem proving & \no{}
	\\
	\hline
	Traoré (2000) \cite{Traore00} & 1.3 & PVS & \cellColorYes{}PVS & \cellColorYes{}model checking / theorem proving & \cellColorYes{}PrUDE
	\\
	\hline
	Kim \etal{} (2002) \cite{KC02} & 1.3 & Z & \no{} & \no{} & \no{}
	\\
	\hline
	Ledang \etal{} (2002) \cite{LS02} & 1.2? & B & \cellColorYes{}B & \cellColorYes{}theorem proving & \no{}
	\\
	\hline
	Balser \etal{} (2004) \cite{BBKRT04} & 1.5 & KIV & \cellColorYes{}KIV & \cellColorYes{}interactive verification & \no{}
	\\
	\hline
	Zhan \etal{} (2004) \cite{ZM04} & 2.0 & Z & \no{} & \no{} & \no{}
	\\
	\hline
	El Miloudi \etal{} (2015) \cite{EE15} & 2.4.1 & Z & \cellColorYes{}Z & \cellColorYes{}consistency checks & \no{}
	\\

	\hline
	\hline
	Knapp \etal{} (2017) \cite{KM17} & 2.5 & interactions & \cellColorYes{}DOL & \cellColorYes{}consistency checks & \no{}
	\\
	\hline
	Rosenberger \etal{} (2021) \cite{RBKR20} & 2.5.1 & interactions & \cellColorYes{}CASL / HETS / SPASS & \no{} & \yes{}
	\\

	\hline

\end{tabular}

	}
	\label{table:translation:tools}
\end{table}

\ifdefined\VersionLong
	\end{landscape}
\fi
\section{Approaches providing operational semantics for UML state machines}\label{sec:formalizing}

Different from the translation-based approaches, the second kind of approaches directly provides an operational semantics to UML state machines, usually by defining inference rules.
Various verification techniques can then be conducted based on the operational semantics.
The benefits of this kind of approaches are:
\begin{enumerate}
	\item they do not rely on the target formal languages, thus no redundancies are introduced, and 
	\item the semantic steps defined in the operational semantics directly coincide with the UML state machine semantic step, \ie{} the run-to-completion step.
\end{enumerate}
Semantic approaches are harder to classify than translation-based approaches.
Therefore, we follow a mostly chronological approach based on the UML version:
we first describe UML~1.x approaches (\cref{ss:semantic:1}),
then the UML~2.0 approaches by Schönborn \etal{} and extensions (\cref{ss:semantic:2}),
then one approach for UML~2.1 (\cref{ss:semantic:2.1})
and finally some approaches for UML~2.4 (\cref{ss:semantic:2.4}).
Finally, we summarize the approaches in \cref{ss:summary}.

\subsection{Operational semantics for UML 1.x}\label{ss:semantic:1}
\paragraph{Approaches using EHAs}
Hierarchical Automata require a strict hierarchical structure. The existence of inter-level transitions and local transitions breaks the hierarchical structure. 
Extended Hierarchical Automata (EHAs) extend Hierarchical Automata to deal with inter-level transitions \ie{} an inter-level
transition which crosses multiple states will be assigned to the outermost Sequence Automaton.
Although approaches using EHAs as an intermediate representation do not ``directly'' provide
the operational semantics, EHAs still resemble UML state machines in the hierarchical structure,
and EHAs are equipped with an operational semantics. %
For this reason we consider this kind of approaches as directly providing operational semantics.
Latella \etal{}~\cite{LMM99} %
are among the pioneers who began to focus on formalizing UML 
statecharts (instead of other variants of statecharts) semantics.
They use a slightly modified variant of EHAs as an intermediate model, and map the UML-statecharts into an EHA.
The  hierarchical structure of UML statecharts and EHAs makes the translation structured and intuitive. 
Then they define the operational semantics for EHAs using Kripke structures. 

A following work by Gnesi \etal{}~\cite{GLM02} %
extends~\cite{LMM99} to include multicharts, \ie{} multiple UML state machines communicating asynchronously, using a non-deterministic choice of event dispatch between the various components.
The work also discusses how to incorporate the semantics into the model checking tool \jack{}~\cite{JACKreport94,GLM99}.
This approach covers a quite restricted subset of UML state machine structures: no pseudostates (except the initial pseudostate) are considered, neither entry/exit/do behaviors nor deferred events are considered, and the triggering 
events are restricted to signal and call events without parameters.
Even though the ultimate goal is to use \jack{} for the automated verification, the authors do not discuss the implementation of the translation, and it is not clear whether it has been done.\label{newtext:GLM02}
In fact, the authors themselves explicitly focus on the design issues rather than on the implementation ones.

Dong \etal{}~\cite{DWQQ01} further extend EHAs to support more features such as entry/exit behaviors or parameters in behaviors, and provide a formal semantics for a subset of UML statecharts based on this EHA model. 
The authors discuss the findings on the cost of solving conflicts introduced by concurrent 
composite states, and the importance of modeling with multiple objects instead of modeling them with concurrent regions within one UML state machine.
They consider the non-determinism caused by multiple concurrent state machines, which was not captured by~\cite{LMM99}.
This work is extended in~\cite{WDQ02}.
Note that slicing has also been discussed, notably in \cite{FL05,Miao11} for statecharts, and in~\cite{ACHHLT13,AAC20} for extended finite state machines (EFSMs).

\paragraph{Other approaches for UML 1.x}

Von der Beek~\cite{VDB02} also formalizes a partial set of UML statecharts, partially based on the work proposed in~\cite{LMM99}.
But it supports some more features compared to~\cite{LMM99}, such as history mechanisms, entry and exit %
	behaviors.
The syntax used in this work is called a UML-statechart term, which is inductively defined on three kinds of terms, \viz{} basic term, or-term and and-term.
All of them contain basic information about a state such as a unique ID, entry and exit %
	behavior, and sub-terms (for or-term and and-term) which contain the hierarchical information of a UML state machine.
UML-statechart terms basically represent static information about UML statechart vertices. 
Inter-level transitions are captured by explicitly specifying source restrictions and target determinators in an or-term; this notation follows the idea of~\cite{LMM99}.

In~\cite{Kwon00}, Kwon proposes another approach using Kripke structures, and aims at model checking UML statecharts.
Similarly to~\cite{VDB02}, Kwon uses terms as the syntax domain of UML statecharts, which represent state hierarchy in the form of subterms as a field in a term.
But~\cite{Kwon00} uses conditional rewriting rules to represent the transition relation in a UML statechart (while \cite{VDB02} explicitly defined five structural operational semantics rules).
Then the semantics of UML statecharts is defined as a Kripke Structure.
\cite{Kwon00} also provides a translation from the defined Kripke structure to the input language of the \smv{} model checker, which we have discussed in \cref{ss:smv}.

Eshuis and Wieringa~\cite{EW00} provide an operational semantics for UML statecharts using LTS.
This work focuses more on the communication and timing aspect of UML statecharts. 
It also considers object construction and destruction, which is not always considered by the other approaches.
Model checking is performed using the ATCTL logics and the Kronos model checker~\cite{Yovine97}, and experiments are tabulated in~\cite{JW02}.

Lilius and Paltor~\cite{LP99,LP99b} provide an abstract syntax and semantics for a subset of UML state machines. 
They use terms as syntax model and consider most features of UML state machines.
Although it does not define a clear semantic model, their work formalizes the run-to-completion step semantics into an
algorithm. The algorithm is given at an abstract level and many concepts such as history pseudostates
and completion events are described in a rather informal manner. But the procedure of the run-to-completion step is 
properly described.
Some features such as join, fork, junction, choice vertices are unspecified and, instead, it is claimed that these pseudostates can be replaced with extra transitions.

Reggio \etal{}~\cite{RACH00} provide UML state machines with a formal semantics given as an LTS.
This work considers an early version (1.3) of UML specifications and discusses some 
inconsistencies and ambiguities in the specification. The work does not provide a clear syntax model 
and UML state machines are not represented formally.
But it does discuss in detail the event dispatching, as well as the way the events are inserted into the queue: notably, the authors assume that ``it is better to have a mechanism ensuring that when two events are received in some order they will be dispatched in the same order'', and therefore the event queue is modeled as a multiset of events---and neither as a queue nor as a set.\label{newtext:RACH00}

Damm \etal{}~\cite{DJPV02} provide a formal semantics for a kernel set of UML in order
to model real-time applications, including static and dynamic aspects of the UML models.
The formalization contains two steps.
Firstly, real-time UML (rtUML) is represented in terms of the introduced concept of ``kernel subset of real-time UML'' (krtUML).
Secondly, krtUML is equipped with a formal semantics.
This approach provides a self-defined action language, which supports object creation/destruction, assignment and operation calls.
The semantics is given in terms of symbolic transition systems, %
a concept originally introduced in~\cite{MP92} as ``synchronous transition systems'', and that can be seen as an extension of transition systems with first-order logic predicates.
UML state machine are just a component of~krtUML, and therefore state machines are not the core of the formalization proposed in~\cite{DJPV02}.
Still, this work provides a good reference for communications between different objects, such as event dispatching and handling.
Additional details are tabulated in~\cite{DJVP03}.

\subsection{Operational semantics for UML~2.0}\label{ss:semantic:2}

\emph{An almost complete operational semantics for UML~2.0}
Schönborn~\cite{Schonborn05} provides in his \emph{Diplomarbeit} (Master thesis) a very comprehensive analysis about UML~2.0 behavioral state 
machines, including discussions about detailed semantics of each feature, and an exhibition of numerous ambiguities in the UML specification.
This approach covers almost all features of UML~2.0 state machines, 
except for choice, termination pseudostates and completion events.
In addition, junction pseudostates are considered as syntactic sugar and are said to be easily represented by separate transitions.
More precisely, Schönborn argues in~\cite{Schonborn05} that junction pseudostates ``are used as a shorthand notation for collections of transitions'' and that ``submachine states (and therefore also entry and exit pseudostates) can be compiled away'' (the UML specification explicitly mentions that a submachine state is ``semantically equivalent to a composite State'' \cite[p.\,311]{UML251}).
Even better, not only most of the syntax is considered in this work, but the author even discards some restrictions from the UML specification, claiming that his semantics is still valid in the absence of these restrictions;
put differently, the syntax considered in~\cite{Schonborn05} can be seen as \emph{larger} than the official specification.

In a first part, a formal syntax for UML state machines is defined (made of a 10-tuple to encode states, regions, substates, behaviors, transitions, etc.).

In a second part, the formal semantics is introduced.
Arguing that ``flattening'' the hierarchical structure may lead to an unnecessary state explosion, the semantics is presented in a hierarchical manner whenever possible.
Many auxiliary functions are defined to capture the execution of a run-to-completion step, such as collecting 
all actions generated during transition execution and putting them in the event pool. 
Priorities are handled in a particularly precise manner.

This work not only can be considered as a very detailed discussion about the semantics of UML~2.0, but it also contributes to the analysis of \emph{ambiguities} in the UML~2.0 specification (see~\cite{FSKdR05}, discussed below).

In~\cite{FKS05}, a formalization very similar to~\cite{Schonborn05} is given.
The relationship between both works is unclear: one author is coauthor of both works, and \cite{Schonborn05} is a Diplomarbeit while \cite{FKS05} is a technical report; none of these two works (both dating from~2005) cite the other one.
In~\cite{FKS05}, again, most syntactic aspects are formalized: final pseudostates, composite states, deferred events, conflicts, priorities, both shallow and deep history pseudostates, entry/exit behaviors, completion, internal transitions, join/fork, and run-to-completion step.
As in~\cite{Schonborn05}, several ambiguities in the UML specification are exhibited.
The selection mechanism of events is not considered in the paper; neither is the execution of actions.
No tool implementation is mentioned.

\emph{Extensions and other works}
In a separate paper~\cite{FSKdR05}, Schönborn and additional co-authors discuss 29 new ``unclarities'' in the UML~2.0 state machine specification.
They can consist of ambiguities, inconsistencies, or unnecessarily strong restrictions;
these unclarities are clearly linked to Schönborn's Diplomarbeit~\cite{Schonborn05}, where he already had spotted such issues, and discussed unnecessarily strong restrictions (actually lifted in his formalization).
The work in~\cite{FSKdR05} is clearly not a formalization of UML state machines, but can help the community to better formalize the specification.
(Also note that a 30th ambiguity is pointed out by~\cite{FS06}.)

Fecher and Schönborn use in~\cite{FS06} ``core state machines'' %
as the semantic domain for UML state machines. 
A core state machine is a 7-tuple including a set of states (including region and parent relations), a set of do actions, a set of deferred events, a set of transitions, an initial state, a set of variables, and an initial variable assignment. 
History is explicitly described by a mapping from a region to its direct substate. 
The work firstly formalizes both 
syntax and semantics of the core state machine. 
This paper considers more UML state machine features.
Although this approach is one of the most complete approaches in term of syntax considered, it suffers from some limitations.
The run-to-completion step of a UML state machines is not properly defined.
The transformation steps from a UML state machine to a core state machine are provided, but the steps are not formally defined:
instead, only natural language descriptions with example illustrations are given.
Moreover, the translation is complex since a lot of auxiliary vertexes need to be added, such as enter/exit vertices.
This is actually obvious in the article own figures, which are barely readable (see, \eg{} \cite[p.\,258]{FS06}).
These limitations may make it difficult for automatic tool development---which indeed does not seem to have been done.

Two other approaches were considered by this group of authors.
First, in~\cite{FKRB06}, Fecher \etal{} are specifically interested in the compositional aspect: they define a compositional operational semantics for ``flat'' UML state machines (with only simple transitions guarded by expressions on variables).

Second, in~\cite{FHSS09}, Fecher \etal{} define a semantics to model persistent nondeterminism, which can model faulty systems.
The paper is specifically interested in refinement, and uses so-called ``$\mu$-automata''.

Finally, Lano and Clark propose in~\cite{LC07} an axiomatic semantics for a subset of the syntax of UML \emph{protocol} state machines, based on~\cite{FSKdR05}.
Although we are here interested in \emph{behavioral} state machines, the authors give enough hints so that its formalization can be directly applied to behavioral state machines---basically without entry/exit/do behaviors.
The supported syntax is reasonably large, including composite states, deferred events, and history states; however, junction, choice, internal transitions, etc., are left out.
They use the B formalism~\cite{Abrial96} as a backend to perform formal verification.

\subsection{Operational semantics for UML~2.1}\label{ss:semantic:2.1}

In~\cite{Seifert08ICFEM}, Seifert proposes a formal semantics for UML state machines, following as much as possible the (informal) semantics, and inspired by existing works~\cite{BBKRT04,LP99,LMM99Spin}.
The full translation is tabulated in~\cite{Seifert08report}.
While this work ultimately aims at generating test cases (which goes beyond the scope of this survey), the formalization of the UML state machine semantics is sufficiently interesting to be included.
The set of syntactic elements covered by the formalization is explicitly stated, and includes notably composite states, and a careful handling of the run-to-completion step; however, more basic features (entry and exit behaviors, forks, joins, …)\ are left out.
A subset of the Java programming language is used to express guard and action (``behaviors'') expressions.
The actual test generation is made using the \textsc{Teager} tool suite~\cite{SS06,SS06report}.

\subsection{Operational semantics for UML~2.4}\label{ss:semantic:2.4}

Liu \etal{}~\cite{LLACSWD13} provide a formal operational semantics for UML state machines using LTS.\footnote{%
	Note that the authors of this survey were involved in~\cite{LLACSWD13}.
}
The approach covers all the features of UML state machines except for time events.
The approach also considers asynchronous/synchronous communications between objects.
\LongVersion{\footnote{%
	In fact, we had started to draft a preliminary version of this survey before defining our semantics in~\cite{LLACSWD13}; hence, our definition of the semantics benefited from the weaknesses and limitations of other approaches described here.
}}

More recently, in~\cite{BTJBD21}, Besnard \etal{} propose an approach to model not only systems, but also properties, in a unified UML framework.
Their approach uses a ``Semantic Transition Relation'' interface, presented with a formal syntax strongly inspired by the Lean theorem prover~\cite{MU21}.
The presentation is not detailed enough to get a full idea of what syntactic elements are considered; but the approach is particularly interesting due to its unified view.
In addition, not only verification, but also online monitoring, can be performed.
These algorithms are implemented in the EMI framework~\cite{BBDJOT17,BBJTD18,BBJTD18MODELS,BTJBD19MODELS}.

\subsection{Summary}\label{ss:summary}
\subsubsection{Summary of features}

We summarize the surveyed semantic-based approaches in \cref{table:semantics:features}.
We use the same conventions as in \cref{table:translation:features}.
In addition, ``$\indirectly{}$'' denotes an indirect handling, \eg{} submachine states are not supported in~\cite{Schonborn05}, but the author reminds that they can be encoded using composite states. %
Also see the threats to validity discussed in \cref{newtext:threats-scores} concerning our method to compute the ``sum'' value for each work.\label{newtext:threats-scores-recall}

First, as in \cref{table:translation:features}, it is clear from \cref{table:semantics:features} that most approaches do not take into consideration many syntactic elements of UML state machines.
Similarly, few such elements are supported by many approaches---with the exception of orthogonal states and inter-level transitions.
In fact, only five approaches support more than half of the syntactic elements (\cite{LP99b,Schonborn05,FKS05,FS06,LLACSWD13}), with two of them (\cite{FS06,LLACSWD13}) supporting a very large majority of elements.

Overall, the semantic-based approaches support in average more elements than the translation-based approaches.
This can be seen as a paradox: indeed, semantic-based approaches are often theoretical, and it can be natural to not consider all elements, if the missing elements can be encoded themselves into supported elements (this could be the case for \eg{} submachine states, junctions, and even history pseudostates).
In contrast, translation-based approaches missing these elements have no other choice but translating them into supported elements, which makes the approach incomplete.

\subsubsection{Tool support}

We review the tool support offered by the translation-based approaches in \cref{table:semantics:tools}.
The mostly theoretical nature of the semantic-based approaches is confirmed by \cref{table:semantics:tools} as most approaches provide strictly no tool support, neither a parser from an existing formalism for UML state machines, nor a model-checker supporting the chosen semantics.
Sadly, out of the two most complete approaches (\cite{FS06,LLACSWD13}), only the latter provides a tool support.
As mentioned earlier, we suspect that the quite complex formalization of~\cite{FS06} made it rather delicate to allow for a practical implementation.

\ifdefined\VersionLong
	\newpage
	\begin{landscape}
\fi

\begin{table}
	\caption{UML state machine features supported by each semantic approach}

	\setlength{\tabcolsep}{\ifdefined\VersionLong2pt\else1pt\fi}
	\scalebox{\ifdefined\VersionLong.75\else.61\fi}{

\begin{tabular}{ |c|c|c|c|c|c|c|c|c|c|c|c|c|c|c|c|c|c|c|c|c|}
	\hline
	\rowcolor{colorHead}
	\bfseries{}Approach & nb & UML & 			\multicolumn{2}{c|}{\cellcolor{colorHead}States} & 			\multicolumn{6}{c|}{\cellcolor{colorHead}Pseudostates} & 			Entry/exit & 			\multicolumn{3}{c|}{\cellcolor{colorHead}Transitions} & 			RTC & 			Variables & 			Deferred & 			Time & 			Multiple & 			Sum 			
	\\
	\rowcolor{colorHead}& cit. & v & 			ortho & subm & 			fk/jn & junct. & choice & sH & dH & en/ex & 			behv & 			intern & interl & compl & 			& 			& 			events & 			& 			charts 			& 		\\
	\hline
	Lilius \etal{} (1999) \cite{LP99b} & 	\cellcolor{blue!26}116 & 1.3 & \yes{} & \no{} & \yes{} & \yes{} & \no{} & \yes{} & \yes{} & \no{} & \yes{} & \yes{} & \no{} & \yes{} & \yes{} & \partially{} & \yes{} & \yes{} & \no{} &\cellcolor{blue!67}11.5
	\\
	\hline
	Reggio \etal{} (2000) \cite{RACH00} & 	\cellcolor{blue!24}108 & 1.3 & \no{} & \no{} & \no{} & \yes{} & \no{} & \no{} & \no{} & \no{} & \no{} & \no{} & \yes{} & \no{} & \no{} & \no{} & \no{} & \no{} & \no{} &\cellcolor{blue!11}2.0
	\\
	\hline
	Eshuis \etal{} (2000) \cite{EW00} & 	\cellcolor{blue!18}81 & 1.3 & \yes{} & \no{} & \no{} & \no{} & \no{} & \no{} & \no{} & \no{} & \yes{} & \no{} & \no{} & \yes{} & \yes{} & \no{} & \no{} & \yes{} & \no{} &\cellcolor{blue!29}5.0
	\\
	\hline
	Dong \etal{} (2001) \cite{DWQQ01} & 	\cellcolor{blue!11}52 & 1.1 & \yes{} & \no{} & \no{} & \no{} & \no{} & \no{} & \no{} & \no{} & \yes{} & \no{} & \yes{} & \yes{} & \yes{} & \yes{} & \no{} & \no{} & \yes{} &\cellcolor{blue!41}7.0
	\\
	\hline
	von der Beeck (2002) \cite{VDB02} & 	\cellcolor{blue!30}136 & 1.4 & \yes{} & \no{} & \no{} & \no{} & \no{} & \yes{} & \yes{} & \no{} & \yes{} & \no{} & \yes{} & \no{} & \no{} & \no{} & \no{} & \no{} & \no{} &\cellcolor{blue!29}5.0
	\\
	\hline
	Gnesi \etal{} (2002) \cite{GLM02} & 	\cellcolor{blue!13}59 & 1.3 & \yes{} & \yes{} & \yes{} & \no{} & \no{} & \no{} & \no{} & \no{} & \no{} & \no{} & \yes{} & \no{} & \no{} & \no{} & \no{} & \no{} & \yes{} &\cellcolor{blue!29}5.0
	\\
	\hline
	Damm \etal{} (2002) \cite{DJPV02} & 	\cellcolor{blue!20}89 & 1.4 & \no{} & \no{} & \no{} & \no{} & \no{} & \no{} & \no{} & \no{} & \no{} & \no{} & \no{} & \yes{} & \yes{} & \yes{} & \no{} & \yes{} & \yes{} &\cellcolor{blue!29}5.0
	\\
	\hline
	Schönborn (2005) \cite{Schonborn05} & 	\cellcolor{blue!2}9 & 2.0 & \yes{} & \indirectly{} & \yes{} & \indirectly{} & \no{} & \yes{} & \yes{} & \indirectly{} & \yes{} & \yes{} & \yes{} & \no{} & \yes{} & \no{} & \yes{} & \no{} & \no{} &\cellcolor{blue!70}12.0
	\\
	\hline
	Fecher \etal{} (2005) \cite{FKS05} & 	\cellcolor{blue!3}15 & 2.0 & \yes{} & \indirectly{} & \yes{} & \indirectly{} & \no{} & \yes{} & \yes{} & \indirectly{} & \yes{} & \yes{} & \yes{} & \yes{} & \yes{} & \no{} & \yes{} & \no{} & \no{} &\cellcolor{blue!76}13.0
	\\
	\hline
	Fecher \etal{} (2006) \cite{FKRB06} & 	\cellcolor{blue!1}7 & 2.0 & \no{} & \no{} & \no{} & \no{} & \no{} & \no{} & \no{} & \no{} & \no{} & \no{} & \no{} & \no{} & \no{} & \yes{} & \no{} & \no{} & \yes{} &\cellcolor{blue!11}2.0
	\\
	\hline
	Fecher \etal{} (2006) \cite{FS06} & 	\cellcolor{blue!8}38 & 2.0 & \yes{} & \yes{} & \yes{} & \yes{} & \yes{} & \yes{} & \yes{} & \yes{} & \no{} & \yes{} & \yes{} & \yes{} & \partially{} & \yes{} & \yes{} & \yes{} & \no{} &\cellcolor{blue!85}14.5
	\\
	\hline
	Lano \etal{} (2007) \cite{LC07} & 	\cellcolor{blue!4}21 & 2.0 & \yes{} & \no{} & \yes{} & \no{} & \no{} & \yes{} & \yes{} & \no{} & \no{} & \no{} & \yes{} & \yes{} & \unclear{} & \yes{} & \yes{} & \unclear{} & \no{} &\cellcolor{blue!50}8.5
	\\
	\hline
	Seifert (2008) \cite{Seifert08report} & 	\cellcolor{blue!1}5 & 2.1.1 & \yes{} & \unclear{} & \no{} & \no{} & \no{} & \no{} & \no{} & \no{} & \no{} & \yes{} & \yes{} & \yes{} & \yes{} & \yes{} & \no{} & \no{} & \yes{} &\cellcolor{blue!42}7.25
	\\
	\hline
	Fecher \etal{} (2009) \cite{FHSS09} & 	\cellcolor{blue!1}7 & 2.0 & \no{} & \no{} & \no{} & \no{} & \yes{} & \no{} & \no{} & \no{} & \no{} & \no{} & \no{} & \no{} & \no{} & \yes{} & \no{} & \no{} & \yes{} &\cellcolor{blue!17}3.0
	\\
	\hline
	Liu \etal{} (2013) \cite{LLACSWD13} & 	\cellcolor{blue!12}54 & 2.4.1 & \yes{} & \yes{} & \yes{} & \yes{} & \yes{} & \yes{} & \yes{} & \yes{} & \yes{} & \yes{} & \yes{} & \yes{} & \yes{} & \yes{} & \yes{} & \no{} & \yes{} &\cellcolor{blue!94}16.0
	\\
	\hline
	Besnard \etal{} (2021) \cite{BTJBD21} & 	\cellcolor{blue!0}0 & 2.5.1 & \no{} & \no{} & \no{} & \no{} & \yes{} & \no{} & \no{} & \no{} & \yes{} & \no{} & \no{} & \no{} & \yes{} & \yes{} & \yes{} & \no{} & \no{} &\cellcolor{blue!29}5.0
	\\
	\hline
	\hline
	\cellcolor{colorHead}Features supported & -  & -  & 11.0 & 5.25 & 7.0 & 6.0 & 4.0 & 7.0 & 7.0 & 4.0 & 8.0 & 6.0 & 10.0 & 9.0 & 9.75 & 9.5 & 7.0 & 4.25 & 7.0 & 17

	\\
	\hline
	\cellcolor{colorHead}\% & -  & -  & \cellcolor{blue!68}68\,\% & \cellcolor{blue!32}32\,\% & \cellcolor{blue!43}43\,\% & \cellcolor{blue!37}37\,\% & \cellcolor{blue!25}25\,\% & \cellcolor{blue!43}43\,\% & \cellcolor{blue!43}43\,\% & \cellcolor{blue!25}25\,\% & \cellcolor{blue!50}50\,\% & \cellcolor{blue!37}37\,\% & \cellcolor{blue!62}62\,\% & \cellcolor{blue!56}56\,\% & \cellcolor{blue!60}60\,\% & \cellcolor{blue!59}59\,\% & \cellcolor{blue!43}43\,\% & \cellcolor{blue!26}26\,\% & \cellcolor{blue!43}43\,\% & 100\,\%

	\\

	\hline

\end{tabular}

	}
	\label{table:semantics:features}
\end{table}

\ifdefined\VersionLong
	\end{landscape}
\fi

\begin{table}
	\caption{Semantic approach: tool support}

	\ifdefined\VersionLong\else
		\scriptsize
	\fi
	\setlength{\tabcolsep}{2pt} %

\begin{tabular}{ |c|c|c|c|c|}
	\hline
	\rowcolor{colorHead}
	\bfseries{}Approach & UML v & Verification tool & Verification & Translation tool
	\\
	\hline
	Lilius \etal{} (1999) \cite{LP99b} & 1.3 & \cellColorYes{}\vuml{} & \cellColorYes{}model checking & \unclear{}
	\\
	\hline
	Reggio \etal{} (2000) \cite{RACH00} & 1.3 & \no{} & \no{} & \no{}
	\\
	\hline
	Eshuis \etal{} (2000) \cite{EW00} & 1.3 & \no{} & \no{} & \no{}
	\\
	\hline
	Dong \etal{} (2001) \cite{DWQQ01} & 1.1 & \no{} & \no{} & \no{}
	\\
	\hline
	von der Beeck (2002) \cite{VDB02} & 1.4 & \no{} & \cellColorYes{}LTL model checking & \no{}
	\\
	\hline
	Gnesi \etal{} (2002) \cite{GLM02} & 1.3 & \cellColorYes{}\jack{} & \cellColorYes{}ACTL model checking & \no{}
	\\
	\hline
	Damm \etal{} (2002) \cite{DJPV02} & 1.4 & \no{} & \no{} & \no{}
	\\
	\hline
	Schönborn (2005) \cite{Schonborn05} & 2.0 & \no{} & \no{} & \no{}
	\\
	\hline
	Fecher \etal{} (2005) \cite{FKS05} & 2.0 & \no{} & \no{} & \no{}
	\\
	\hline
	Fecher \etal{} (2006) \cite{FKRB06} & 2.0 & \no{} & \no{} & \no{}
	\\
	\hline
	Fecher \etal{} (2006) \cite{FS06} & 2.0 & \no{} & \no{} & \no{}
	\\
	\hline
	Lano \etal{} (2007) \cite{LC07} & 2.0 & \cellColorYes{}B & \cellColorYes{}consistency & \no{}
	\\
	\hline
	Seifert (2008) \cite{Seifert08report} & 2.1.1 & \cellColorYes{}Teager & \cellColorYes{}test generation & \yes{}
	\\
	\hline
	Fecher \etal{} (2009) \cite{FHSS09} & 2.0 & \no{} & \no{} & \no{}
	\\
	\hline
	Liu \etal{} (2013) \cite{LLACSWD13} & 2.4.1 & \cellColorYes{}\usmmc{} & \cellColorYes{}model checking & \yes{}
	\\
	\hline
	Besnard \etal{} (2021) \cite{BTJBD21} & 2.5.1 & \cellColorYes{}\EMI{} & \cellColorYes{}MC and monitoring & \yes{}
	\\

	\hline

\end{tabular}

	\label{table:semantics:tools}
\end{table}

\section{Tool support}\label{section:tools}

In this section, we discuss tool support for verifying UML state machines.
There are both commercial and academic tool supports for UML modeling.
Commercial tools include notably
	Eclipse Papyrus\footnote{\url{https://www.eclipse.org/papyrus/}},
	IAR Visual State\footnote{\url{https://www.iar.com/products/iar-visual-state/}},
	IBM Rhapsody\footnote{\url{https://www.ibm.com/fr-fr/products/uml-tools}},
	Microsoft Visual Modeler,
	or
	Yakindu\footnote{\url{https://www.itemis.com/en/yakindu/state-machine/}}.
	
Also, some open-source tools take as input UML state machines, notably PlantUML\footnote{\url{https://plantuml.com/}}
and VUML\footnote{\url{https://sourceforge.net/projects/vuml/}}; however, these two tools are purely syntactic, and no analysis (simulation, verification) is possible.
In addition, we noticed a number of shortcomings in PlantUML, notably with orthogonal composite states (impossibility to draw ``cross-border'' transitions originating from or targeting a state belonging to a region of an orthogonal composite state; or define an entry point on an orthogonal composite state); in other words, even such a purely syntactic tool does not support the full UML specification.
In addition, Umple~\cite{LFBBGA+21} allows some automated code generation from UML state machines to a number of languages (including Java).

To the best of our knowledge, current non-academic commercial tools only support the design/graphical editing of UML models, or perform some limited ``verification'', but without any publicly available academic foundation---hence, we discard them in our survey.

Some academic prototype tools were developed based on either the translation or the semantic approaches, and aim at automatically verifying UML state machines.
We survey these tools in \cref{ss:tools} and draw comparisons and conclusions in \cref{ss:tools:conclusions}.

\subsection{Surveying tools for verifying UML state machines}\label{ss:tools}

\paragraph{\vuml{}}
\vuml{}~\cite{LPtool99} aims at automatically verifying UML model behaviors specified by UML statechart diagrams.
This tool utilizes \spin{} as a backend to perform model 
checking and creates a UML sequence diagram according to the counterexample provided by \spin{}. The 
formal semantics is defined in~\cite{LP99b}.
The authors also conduct a case study  with the production cell example in~\cite{LP99}.

\vuml{} aims at checking collaborations of UML models instead of a single UML state machine.
\vuml{} provides an event generator to emulate external events without 
parameters %
and removes external events carrying parameters in order to avoid state space explosion. 

\vuml{} can check the following properties: deadlock, livelock, reaching an invalid state, violating a constraint on an object, sending an event to a terminated object, overrunning the input queue of an object, and overrunning the deferred event queue.

In order to verify LTL formulas with UML, the user needs to understand the \promela{} model to come up with a proper LTL formula, which we consider to be a potential drawback.

A nice advantage of \vuml{} is that, when the property is violated, it converts the counterexample output by \spin{} into a UML sequence diagram, hence offering a visual trace to the designer.

\vuml{} does not seem to be either available online nor maintained anymore.\footnote{%
	The page cited in~\cite{LPtool99} (\url{http://www.abo.fi/~iporres/vUML/vUML.html}) does not seem to be anymore available.
	Also note that \vuml{} should not be confused with VUML (standing for Visual UML), an open source project\FinalVersion{ available} at \url{https://sourceforge.net/projects/vuml/}.
}

\paragraph{\jack{}}
Gnesi \etal{}~\cite{GLM99} provide an algorithm to support direct model checking UML statecharts based on 
the formal semantics they have defined in~\cite{LMM99}.
The implementation is based on the tool set \jack{} (``Just Another Concurrency Kit'')~\cite{BGL94}, which is an environment based on %
process algebras, automata and a temporal logic formalism.
Different components of the \jack{} tool set communicate with the FC2 format.
There is a model checking tool in the \jack{} tool set named AMC, which supports ACTL model checking.
The system should be translated into the FC2 format first in order to utilize the AMC component.
The users also need to specify their own ACTL property according to the model.
This requires users to have a knowledge of model checking, the underlying model as well as temporal logic formulas.

\paragraph{\hugo{}}
Knapp \etal{}~\cite{SKM01} developed a tool called \hugo{}, that translates UML state machines into \promela{}, that is then verified using the \spin{} model checker.
\hugo{} requires the presence of Java and (of course) \spin{} to be executed.
\hugo{} used to be available online (see \cref{tab:tools:availability}) in the form of a binary; no source code is available.

\paragraph{\hugort{}}\label{paragraph:HugoRT}
\hugort{} is a UML model translator for model checking and code generation.
The current \hugort{} is a rewrite of both \hugo{}~\cite{SKM01} and \hugort{}~\cite{KMR02}.
A UML model containing active classes with state machines, collaborations, interactions, and OCL constraints can be translated into Java, C++, Arduino, \promela{} (\spin), and timed automata (\uppaal{}), by first representing UML state machines into an intermediate common language called Smile.
Several analyses based on partial evaluation are used on the Smile-level to produce performant and readable code.
A similar approach is used for an intermediate representation of UML interactions in the language Ida, which can be translated to \promela{} and timed automata.

As of today (2022), all language constructs for state machines are supported by \hugort{}, with the exception of submachines, connection point references, and entry/exit points.

\hugort{} requires a Java environment to work, but is then multi-platform for the same reason.
\hugort{} is available online for download (see \cref{tab:tools:availability}) in the form of a binary; the license is unclear, but the tool's Web page invites interested users to write to an email address to obtain the source code.
Several applications were made, notably to model checking (possibly timed) interactions~\cite{KW06,SK16}, and to coloring test cases for software product lines~\cite{KRS14}.
The development is still active as of~2022.

\paragraph{ASM-based Verification Tool}
Shen \etal{}~\cite{SCH02} introduce a tool based on an ASM model checker (which is itself based on the \smv{} model checker). 
The semantics they adopt is defined in~\cite{CGHS00}.
This tool set supports both static and dynamic checks of a UML diagrams.
For static aspects, syntax as well as well-formedness rules given by OCL can be checked. Static views in UML, such as object diagram and class diagram can also be
transformed into ASM and checked. For the dynamic aspects, UML state machine diagrams are transformed into ASM models, and an ASM model checker is invoked to do the model checking. 
This tool takes UML diagrams specified in the XMI format as input and outputs a counterexample in the form given by the \smv{} model 
checker (since the dynamic checking component of this toolset is based on the \smv{} model checker). The 
counterexample trace can be fed to their analysis tool, which will analyze the error trace and 
produce some UML diagrams such as sequence diagrams or a collaboration diagram to the users.
Details about the tool are described in~\cite{CGHS00} and details about the
transformation procedures are discussed in~\cite{BCR00}.

\paragraph{\tabu{}}
Beato \etal{}~\cite{BBCD05} introduce a tool called \tabu{} (``Tool for the Active Behavior of UML'').
\tabu{} takes UML diagrams (activity and state diagrams) in the form of XMI as input, automatically translates them into an \texttt{.smv} representation (the input format of \smv{}) and calls the Cadence \smv{} model checker to verify the UML model.
In addition, \tabu{} also provides an assistant for writing LTL/CTL properties to verify against the model.
This feature  makes the underlying model and the translation procedure transparent to the users, and solves the problem faced by \vuml{}~\cite{LP99b} to some extent.

The translation covers most UML~2.0 features (though not described in detail in their paper) except for synchronization states, events with parameters and dynamic creation 
and destruction of objects. It also provides guides in writing properties. 
A limitation is that the counterexample is given in the input format of \smv{}, which is not intuitive for model designers to map to their models.

\paragraph{\proco{}}
\proco{} translates a UML state machine (version~1.4, described in XMI format supported by the CORA tool~\cite{AP04}) into \promela{}, which is discussed in~\cite{JDJLP06}.
No details are discussed in that paper, but the paper reports the bugs found by the tool, which show its practical effectiveness.
\paragraph{UML-B State Machine Animation Tool}
UML-B state machine Animation~\cite{SSB10rep} is able to translate a UML-B diagram into an Event-B representation, and utilizes ProB~\cite{ProB}---a plug-in of the Rodin platform~\cite{ABHHMV10}---to perform the simulation and model checking tasks.
The tool is still available: it can be installed from the Rodin platform, and is still maintained (latest version is~3.7 as of June~2022).

\paragraph{\usmmc{}}
Liu \etal{} implemented into the \usmmc{} tool~\cite{LLSZWD13} the operational semantics defined in~\cite{LLACSWD13}.\footnote{%
	Note that some of the authors of this survey have been involved in the development of \usmmc{}.
}
The tool supports most features of UML state machines %
and is capable of model checking various properties, such as safety, deadlock-freeness, and LTL.
Although the verification of more elaborated properties (written in LTL) requires the mastering of temporal logics, model checking safety properties as well as deadlock-freeness and liveness can be used without any additional knowledge.
\usmmc{} is a standalone tool using the \pat{} model checking library~\cite{SLDP09}.

\paragraph{\EMI{}}
In~\cite{BTJBD21}, the tool \EMI{} (EMI stands for ``embedded model interpreter'') \cite{BBDJOT17,BBJTD18,BBJTD18MODELS} is used to perform monitoring and formal verification of UML state machines.
The tool is equipped with a well-founded theory, described in several publications.
LTL model checking and deadlock-freeness checking are both supported.
The tool is primarily available for Linux, with runtime execution of UML models available for Windows and MacOS too, but not formal verification.

The underlying model-checker is OBP2, which can be interfaced with either \EMI{} or \AnimUML{} (see below), to perform formal verification on the UML model executed by \EMI{} or \AnimUML{}.
The authors defined some generic interfaces between OBP2 and EMI execution engines such that they can use other model-checkers (tested with a FPGA-based model-cheker called Dolmen~\cite{FTL22}) or other controllers.
\EMI{} has not been officially released neither put in open-source yet.
The tool is still in active development as of~2022.

\paragraph{\AnimUML{}}
Recently, a tool called \AnimUML{} was proposed by Jouault \etal{} in~\cite{JBLTBD20,JSBLTBD21}, and uses the same model checker (OBP2) as \EMI{}.
One of its specificities is to allow for \emph{partial} UML models, and to maintain a list of ``semantic relaxation points'', as the UML specification is sometimes ambiguous.
The set of syntactic features is rather small, but additional diagrams such as sequence diagrams can also be considered.
\AnimUML{} also has the possibility to represent counterexamples as UML sequence diagrams.

\paragraph{KandISTI/UMC toolset}
As mentioned in \cref{sss:exclusion}, the KandISTI/UMC toolset relies on the UMC formalism, together with properties specified using~UCTL~\cite{BFGM11}.
While UMC is not strictly speaking UML, both formalisms share many similarities, and we believe that KandISTI/UMC could be used to verify UML models, possibly after some manual adaptation or automated translation.
The tool is available for Linux, MacOS and Windows, as well as using a browser equipped with HTML and javascript.
Since it does not strictly speaking take UML state machines as input, we do not review the KandISTI/UMC toolset in \cref{tab:tool}.
\label{newtext:KandISTI}

\subsection{Summary}\label{ss:tools:conclusions}

\newcommand{\cellOpen}{\cellcolor{blue!50!white}Open source}
\newcommand{\cellUnclear}{\cellcolor{orange}Unclear}
\newcommand{\cellClosed}{\cellcolor{yellow}Free (proprietary)}
\newcommand{\cellFree}{\cellcolor{yellow}Free}
\newcommand{\cellNA}{\cellcolor{red}Not available}

\begin{table}
	\caption{Summary of the tools: Features}

	\setlength{\tabcolsep}{2pt} %
	\ifdefined\VersionLong
		\footnotesize
	\else
		\scriptsize
	\fi
\begin{tabular}{|c|c|c|c|c|c|c|c|c|c|}
		\hline 
		\rowcolor{colorHead}
		Tool & Reference & Model checker & GUI & Manual & Reach & DLF & LTL & CTL & Counterexample \\
		\hline
		\vuml{} & \cite{LPtool99} & \spin{} & \noUnsure & \noUnsure & \yes & \yes & \yesUnsure & \noUnsure & \yes \\
		\hline
		\jack{} & \cite{GLM99} & AMC & \noUnsure & \noUnsure & \yes{} & \noUnsure & \noUnsure & \yes{} & \noUnsure \\
		\hline
		\hugo{} & \cite{SKM01} & \spin{} & \no{} & \noUnsure{} & \yes{} & \yes{} & \noUnsure{}  & \noUnsure{} & \unclear{} \\
		\hline
		\hugort{} & \cite{KMR02} & \spin{}/\uppaal{} & \no{} & \yes{} & \yes{} & \yes{} & \yes{} & \yes{} & \yes{} \\
		\hline
		ASM-based & \cite{SCH02} & \smv{} & \noUnsure{} & \noUnsure{} & \yes{} & \noUnsure{} & \noUnsure{} & \noUnsure{} & \yes{} \\
		\hline
		\tabu{} & \cite{BBCD05} & \smv{} & \no{} & \noUnsure{} & \yes{} & \noUnsure{} & \yes{} & \yes{} & \partially{} \\
		\hline
		\proco{} & \cite{JDJLP06} & \spin{} & \noUnsure{} & \noUnsure{} & \yes{}  & \yes{} & \noUnsure{} & \noUnsure{} & \noUnsure{} \\
		\hline
		UML-B & \cite{SSB10rep} & ProB & \yes{} & \yes{} & \yes{} & \noUnsure{} & \noUnsure{} & \noUnsure{} & \noUnsure{}\\
		\hline
		\usmmc{} & \cite{LLSZWD13} & (standalone) & \yes{} & \noUnsure{} & \yes{} & \yes{} & \yes{} & \no{} & \partially{}\\
		\hline
		\AnimUML{} & \cite{JSBLTBD21} & OBP2 & \yes{} & \no{} & \yes{} & \yes{} & \yes{} & \no{} & \yes{}\\
		\hline
		\EMI{} & \cite{BTJBD21} & OBP2 & \yes{} & \yes{} & \yes{} & \yes{} & \yes{} & \no{} & \yes{}\\
		\hline
	\end{tabular}
	\label{tab:tool}
\end{table}

\cref{tab:tool} summarizes some information of the surveyed tools with, from left to right, the tool name and its main reference, the underlying model checking engine (if any), whether the tool features a graphical user interface, whether a decent user manual is publicly available, and the types of properties available: reachability or safety, deadlock-freeness, LTL model checking, CTL model checking, and whether a counterexample can be returned (\partially{} denotes a counterexample in the translated formalism, while \yes{} denotes a counterexample back to the original UML model).
Most of these information were obtained from the associated tool papers and/or documentation, since most tools are not available anymore (see \cref{tab:tools:availability}).
This also explains the large number of ``?'' in \cref{tab:tool}.

\begin{table}
	\caption{Summary of the tools: Availability and platforms}
\setlength{\tabcolsep}{2pt} %
\ifdefined\VersionLong
	\footnotesize
\else
	\scriptsize
\fi
	\scalebox{.9}{
	\begin{tabular}{|c|c|c|c|c|c|c|c|c|}
		\hline 
		\rowcolor{colorHead}
		Tool & URL & Available & Linux & Mac OS & Windows & Started & Latest & License \\
		\hline
		\vuml{} & {\tiny\url{http://www.abo.fi/~iporres/vUML}} & $\no$ & \unclear{} & \unclear{} & \unclear{} & 1999 & 1999? & \cellOpen{} \\
		\hline
		\jack{} & $\no$ & $\no$ & \unclear{} & \unclear{} & \unclear{} & 1994 & 1999? & \cellNA \\
		\hline
		\hugo{} & {\tiny\url{http://www.pst.ifi.lmu.de/projekte/hugo/usage.html}} & \no{} & $\yes$ & \noUnsure{} & \noUnsure{} & ? & 2002 & \cellUnclear{} \\
		\hline
		\hugort{} &
			\scalebox{.7}{{\tiny\url{https://www.uni-augsburg.de/en/fakultaet/fai/informatik/prof/swtsse/hugo-rt/}}}
		& \yes{} & \yes{} & \yes{} & \yesMono{} & 2004 & 2022 & \cellFree{}\\
		\hline
		ASM-based & \no{} & \no{} & \unclear{} & \unclear{} & \unclear{} & 2001 & 2002? & \cellNA\\
		\hline
		\tabu{} & \no{} & \no{} & \unclear{} & \unclear{} & \unclear{} & 2004 & 2005? & \cellNA{} \\
		\hline
		\proco{} & {\tiny\url{http://www.tcs.hut.fi/SMUML/}} & \no{} & \unclear{} & \unclear{} & \unclear{} & 2006? & 2006? & \cellNA{}\\
		\hline
		UML-B & \scalebox{1}{\tiny\url{%
%
				https://www.uml-b.org/
%
		}} & \yes{} & \yes{} & \yes{} & \yes{} & 2010? & 2022 & \cellNA{} \\
		\hline
		\usmmc{} & {\tiny\url{http://www.comp.nus.edu.sg/~lius87/UMLSM.html}} & \no{} &  $\yesMono$ & $\yesMono$ & $\yes$ & 2013 & 2013 & \cellClosed{} \\
		\hline
		\AnimUML{} & {\tiny\url{https://github.com/fjouault/AnimUML}} & \yes{} &  \yes{} & \yes{} & $\yes$ & 2020 & 2022 & \cellOpen{} \\
		\hline
		\EMI{} & {\tiny\url{http://www.obpcdl.org/bare-metal-uml/}} & \yes{} & \yes{} & \indirectly{} & \indirectly{} & 2017 & 2022 & \cellFree{} \\
		\hline
	\end{tabular}
	}
	\label{tab:tools:availability}
\end{table}

We give in \cref{tab:tools:availability} the most recent known URL of each tool, followed from left to right by the actual availability (for download, on the Web), the supported operating systems ($\yesMono$ denotes a support of Linux or Mac~OS, using the \href{https://www.mono-project.com/}{mono} utility, which is an open source implementation of Microsoft's .NET Framework, or for Windows using Cygwin), the earliest and latest modification, and the license (when known).

We draw some conclusions on the surveyed tools in the following.

\paragraph{Delegation to external model checkers}
We notice from \cref{tab:tool} that all the available tools, with the notable exception of \usmmc{} and \EMI{}, just provide a front-end supporting translation  from UML state machines to languages of existing model checkers.
Such a translation may introduce extra cost for the verification procedure.

\paragraph{Soundness}
For the tools using a translation to an external language or engine, the informal translation procedure does not in general guarantee the soundness of the obtained model.
In addition, almost none of the model checkers are ``certified'', in the sense of a fully proved translation and verification.
One may argue that this might be less critical for UML diagrams than for, \eg{} code to be embedded into a safety-critical system; still, this is unfortunate that almost no attempt was performed to provide users with a certified model checker addressing the verification of UML state machines.
The only exception is in~\cite{MCSK17}; but this is not strictly speaking a certified model checking engine, but rather a verified translation (using Isabelle/HOL).

\paragraph{Counterexample}
In most tools, it is hard to map the found vulnerabilities back to the original UML model.
That is, only three tools (\vuml{}, ``ASM-based'' and \hugort{}) map the possible counterexample back to the original UML model.
Other tools either show no counterexample, or exhibit it, but fail in mapping it to the original model (\eg{} \usmmc{}).

\paragraph{Long term availability}\label{newtext:longterm}
An obvious conclusion from \cref{tab:tools:availability} is that, with four notable exceptions (\hugort{}, UML-B, \AnimUML{} and \EMI{}), all tools seem to be unavailable publicly nowadays.
None of the advertised URLs in the papers work nowadays.
Research using search engines all failed, and no code or binary seems to be archived on publicly available code repositories (such as \href{https://github.com/}{GitHub}), nor on long-term archiving venues (such as \href{https://zenodo.org/}{Zenodo}).
This is a major issue of the research related to (formal) verification of UML state machines: most prototypes have been designed in the framework of an academic research, in the context of one (or several) particular paper(s) and the authors did not perform the necessary steps to make their tool publicly available, nor ensuring the long-term conservation of it.
This trend goes by far beyond the context of UML verification, and concerns unfortunately most areas of academic research concerned with tool development.
The fact that most of the surveyed prototypes are relatively old (with three exceptions, the development of all tools surveyed in \cref{tab:tool,tab:tools:availability} started before~2010) does not improve the situation as, at that time, researchers may have been less concerned with long-term availability and experiments reproducibility than nowadays.
Note that even one of the most recent tools (\usmmc{}, developed in~2013) seems to have been lost as well.\footnote{%
	The story of this loss is unfortunately classical: the developer of \usmmc{}
	wrote the tool code during her Ph.D.\ thesis.
	After graduation, her university deleted all her data and, while she did perform some backup on external drives, these were eventually lost over the years.
}

\section{Related surveys} \label{sec:related}

We survey here the previous attempts to summarize the formalization of UML state machines.
In the past\LongVersion{ twenty years}, \LongVersion{and }to the best of our knowledge, there were three main surveys on the subject~\cite{BR04,CD05,LRS10}.

\paragraph{Bhaduri \etal{}}
The first survey \LongVersion{on UML state machines }dates from~2004.
Bhaduri \etal{}~\cite{BR04} summarize approaches which translated variants of statecharts (including STATEMATE statecharts and UML statecharts) into the input language of the \smv{} and \spin{} model checkers.
The survey only covers a subset of approaches, with detailed descriptions and discussions about each individual work.
\LongVersion{%
	The paper also discusses possible future research directions, such as adopting slicing, abstraction-based approaches or compositional verification approaches to reduce the state space in model checking, or exploring abstractions on environments.
}
The paper only lists a subset of works translating statecharts into \smv{} or \spin{}.
It does not provide any comparison among those works or conclusive comments on those works.
Finally, \cite{BR04} focuses on many variants of Harel's statechart~\cite{Harel87,HN96,HLNPPSST90}, such as RSML or UML. 
\LongVersion{%
(A comparison of the semantic differences of these variants can be found in~\cite{CD07}.)
We rather focus specifically here on UML state machines, which is the object-oriented variant of Harel's statecharts.
}

\paragraph{Crane and Dingel}
Crane and Dingel~\cite{CD05} provide in~2005 a categorization and comparison of 26 different approaches of formalizing UML state machine semantics (including denotational and operational semantics).
They categorize those approaches based on the underlying formalism used, and conduct comparisons on other dimensions such as UML state machine features coverage or tool support.
That paper provides a high-level comparison and discussion on different aspects of the existing 26~approaches.
Although the amount of work discussed is not large, it covers different kinds of approaches and provides a good way to categorize those approaches.
The categorization of the works is wide, but the coverage within each categorization is too narrow, as several translation-based approaches are not included.
\paragraph{Lund \etal{}}
In 2010, Lund \etal{}~\cite{LRS10} survey existing works on formalizing UML sequence diagram-like and state machine diagram-like semantics.
Their survey does not focus on a thorough coverage of all the existing approaches; instead, it selectively discusses some representative approaches which fall in one of their categorization criteria. 
\LongVersion{%
There are two dimensions of categorization.
The main dimension is the style of the semantics, namely denotational semantics and operational semantics.
The second dimension captures features such as real-time or probability.
}%
The focus is on both sequence diagrams and state machine diagrams. 
Although the survey provides new comparison criteria such as the supported properties or refinement support, it covers a limited number of approaches only, especially for the UML state machine part. 
\LongVersion{%
	Moreover, formal verification tools for UML state machines are not discussed.
}

\paragraph{Other surveys}

In~\cite{Beeck94}, von der Beeck compares variants of the ``statecharts'' (prior to their formalization by the OMG), notably in terms of concurrency model (true concurrency or interleaving), timeout on trigger transitions, etc.

In~\cite{LM96}, Lee and Yannakakis discuss \emph{testing} finite state machines.

In~\cite{BS01}, Balsamo and Simeoni survey the transformation of UML models into performance models.
This survey came at the early stage of UML formalization, and does not focus only on state machine diagrams, but also considers class, use case, activity, sequence, collaboration, deployment and component diagrams.
It remains therefore relatively shallow on that particular aspect.

In~\cite{MPT03}, two different interpretations of the statecharts step semantics are considered (notably \wrt{} simulation, %
ready trace preorder, failure preorder and trace preorder).

In~\cite{BMP12}, dependability modeling with UML is surveyed.
That survey does not focus specifically on UML state machines.

In~\cite{KM18}, a survey is made on 57 approaches related to the consistency of multi-view models in UML/OCL.
All 14 different UML diagram types are considered (not only state machines); but few approaches cover many diagram types; in fact, even the most comprehensive surveyed approach covers (partially) only five UML diagram types.

Other surveys were proposed for other UML diagrams, \eg{} sequence diagrams in~\cite{MW11}.

\section{Conclusion}\label{section:conclusion}

In this manuscript, we provide a detailed survey of approaches aiming at giving a formal semantics to UML behavioral state machines, thus enabling their automated verification.
We categorize the approaches into two major groups, \viz{} the translation approaches and those directly providing operational semantics.
In each group, we also provide comparisons of the surveyed approaches on dimensions such as UML version, coverage of syntactic features, and tool support.
We also try to provide a focus on tool support and implementation.

\subsection{Main conclusions}\label{ss:conclusions}

\paragraph{Completeness}
In translation-based approaches, numerous works support various syntactic features, but all fail in supporting a complete or near-to-complete subset of the UML syntax.
\cref{table:translation:features} (page~\pageref{table:translation:features}), that summarizes these approaches, shows that no work supports more than 65\,\% of the existing syntactic features, and most works support less than~50\,\%, sometimes far less.
An additional difficulty comes from the fact that, while a certain level of formalization would be expected from works aiming at \emph{formalizing} UML, several works describe their formalization in a rather ``informal'' (textual) manner; this results in numerous ``$\unclear$'' or $\partially{}$ in \cref{table:translation:features}.
However, all syntactic features are supported by at least one translation approach.

Approaches providing UML state machines with a dedicated operational semantics perform better: while many approaches support only a restricted subset of the UML syntax, two (\viz{} \cite{FS06,LLACSWD13}) support almost completely the UML syntax, as shown in \cref{table:semantics:features} (page~\pageref{table:semantics:features}).

These comments on the completeness also raise questions on the usefulness of the UML syntax: since most works (especially translation-based) only support a restricted subset of the UML syntax, are the remaining syntactic features useful?
In other words, are the infrequent syntactic constructs not considered in most papers because they are deemed of low usefulness, or because they lead to complex translations?
The answer probably goes beyond this survey, and it would be worth investigating the actual use of the UML in practice, so as to survey which syntactic features are most commonly used.

\paragraph{Tools long term availability}
A frustrating outcome from \cref{tab:tools:availability} (page~\pageref{tab:tools:availability}) is that most tools developed over the years for the formal verification of UML state machines are now lost (\ie{} they seem completely unavailable online).
While this phenomenon certainly goes beyond this particular research on formalization of UML, it is particularly obvious, as all the tools we surveyed---with only four exceptions---are lost.
This should be in itself a motivation for researchers to handle with much more care the prototypes they develop to ensure they remain permanently available for several reasons:
\begin{oneenumerate}%
	\item public availability;
	\item surveying purpose;
	\item avoiding to re-invent previously coded translations; and,
	\item experiments reproducibility and comparisons.
\end{oneenumerate}%
We believe long-term archiving venues (such as \href{https://zenodo.org/}{Zenodo}, providing a digital object identifier (DOI) to software and data) should be frequently used.
Also, the Software Heritage initiative~\cite{DZ17} can be used to that purpose.\footnote{%
	In addition to manual archiving requests, the \href{https://www.softwareheritage.org/}{Software Heritage initiative} automatically browses, archives and replicates several existing repositories, such as most of the software code stored publicly in large repositories such as \href{https://bitbucket.org/}{BitBucket}, \href{https://github.com/}{GitHub}, \href{https://gitlab.com}{GitLab} or \href{https://hal.archives-ouvertes.fr/}{HAL}.
}

\subsection{Perspectives}\label{ss:perspectives}

\paragraph{Consistency}
An outcome of our survey is that, overall, few works considered large subsets of the UML state machine syntax in their formalization.
Similarly, no single syntactic element surveyed in \cref{table:translation:features,table:semantics:features} was addressed by all works (note that our tables exclude ``trivial'' elements, such as simple states or initial states).
Therefore, a first perspective includes studying the consistency of these works over the formalized elements of the UML syntax.
Whether a common understanding of these elements throughout the literature has been met is not clear, and this understanding could be put in perspective with fUML and SysML (see below).

\paragraph{A full formalization}
Clearly, the perfect formalization of UML state machine diagrams remains to be done, as none of the existing approaches is entirely satisfactory.
One may wonder why it has not been done before.

On the one hand, academic papers may not always aim at completeness, but rather at formalizing some yet uncovered syntactic feature, or showing that (some of) the UML syntax can be encoded using a formalism that was not used for that purpose yet.
Difficulties can also come from the target formalism, \eg{} it can be difficult to encode the complex hierarchy of entry and exit behaviors using a formalism such as Petri nets.

On the other hand, one may wonder whether the entire syntax of UML is useful for practical purposes (see discussion above).
In addition, the OMG semi-formal semantics is intrinsically not (entirely) formal, and obstacles to the formalization can also come from some perhaps unnecessarily complex syntactic elements.

\paragraph{A minimal syntax}
Whether the syntax of UML state machines as described in the specification can be reduced to a minimal subset remains blur.
In other words, can we (really) encode perfectly some constructs into others?

On the one hand, it is for example claimed that a submachine state is ``semantically equivalent to a composite State'' \cite[p.\,311]{UML251}.
Or that ``an entry point is equivalent to a junction pseudostate (fork in cases where the composite state is orthogonal)'' \cite[p.\,311]{UML251}.
We may wonder whether this is perfectly accurate.
For example, if a model contains recursive references (\eg{} a state machine that would contain a submachine state linking to itself), the model becomes ill-formed (infinitely nested); this cannot happen by using only regular composite states, and therefore the two constructs are not strictly speaking equivalent.

On the other hand, it may be relatively straightforward to show that junctions could be encoded using a number of independent transitions.

\paragraph{Adoption of the UML in the industry}
While many academic works targeted the formalization of UML diagrams (as it is shown by this manuscript, dozens of academic works targeted the formalization of the sole UML state machines), the use of UML diagrams in an industrial context is debatable.

First, it is noted in~\cite{BBBKP11} that ``there are few studies of adoption [of the UML] and use in the field''.
Among the 49 papers surveyed in~\cite{BBBKP11}, only 2 aim at surveying the \emph{adoption} of the UML.

Then, in~\cite{Petre13}, Marian Petre reports on 50 interviews with software engineering professionals over 2~years; among these 50 interviews, 35 did not use UML at all, and only 3 out of~50 used it for automated code generation---which can be seen as one of the goals (though not the only one) of using UML~diagrams.
	In addition, among the 11 out of~50 professionals using a selective subset of the UML, only 3 out of~11 used state machine diagrams.
	Petre therefore concludes that, while UML might still be a ``\emph{de facto} standard'', it is ``by no means universally adopted''.
In addition, according to the author~\cite{Petre14}, most reactions from software professionals to this study were that it came with absolutely no surprise (``the response from software professionals was largely `No shit'.''~\cite{Petre14}).

These works show that, despite an extensive academic literature on formalizing (or using) UML diagrams, its adoption in the industry can be largely improved.
A main challenge is not only to understand the gap between the works done by the academic (and notably the formalization works) on the one hand, and the actual need in the software industry on the other hand.\label{newtext:adoption}

\paragraph{Other UML diagrams}
Verification of UML state machines was in most (but not) all cases performed in isolation, while a verification taking other UML diagrams into consideration would be highly welcome.
UML state machines can be considered jointly with sequence diagrams, or class diagrams, or even activity diagrams.
A future global approach supporting various types of diagrams simultaneously would therefore be most interesting.

\paragraph{Extensions}
Few works studied quantitative extensions of state machines with time and probabilities.
Time is briefly mentioned in the official (semi-formal) semantics~\cite{UML251}, while probabilities are not.
But either timing or probabilistic aspects are considered by a minor number of works (notably \cite{Traore00,KMR02,DMY02,TZ05,CJ09,KST17}).
A probabilistic extension of UML statecharts was also proposed in~\cite{JHK02}.
This seems an interesting direction of research to us, as these quantitative extensions can be of great use for practical systems.
Also, real-time extensions of UML were proposed, which includes krtUML~\cite{DJPV05}, but also the MARTE profile~\cite{MA09}.
An orthogonal question is also whether a real-time semantics is necessary for the UML, or if a discrete semantics can be sufficient.\label{newtext:realtimeVSdiscrete}

Another interesting perspective is the relationship with the Foundational UML (fUML)~\cite{fUML15}, which aims at providing an executable semantics for a subset of the UML syntax.
Notably, is the semantics considered in the works surveyed in this manuscript compatible with that of fUML?
Is the subset of syntactic features of UML state machines that do not belong to fUML useful for practical needs?
This would deserve a survey in itself.\label{newtext:conclusion:fUML}
A similar perspective can be made on SysML.

\paragraph{Tool support}
As surveyed in \cref{tab:tool} (page~\pageref{tab:tool}), the tool support for formal UML verification remains unsatisfactory.
Beyond the fact that most tools are now unavailable (see discussion above), none of the tools is entirely satisfactory: the subset of the considered syntax is usually small, counterexamples are rarely mapped back to the original models, and most tools are prototypes with no or basic GUI, and few OS support (not even mentioning usage through a Web access).
This is probably more an engineering issue rather than a research issue, considering the fact that a satisfactory operational semantics was defined in at least two works (\cite{FS06,LLACSWD13}).

It remains unfortunate that no team dedicated enough efforts into providing the community with a decent verification engine for UML state machines; nevertheless, three tools still under development as of~2022 (\hugort{}, \AnimUML{} and \EMI{}) do propose a satisfactory support for UML state machines verification.

Finally, a further perspective would be to prove the correctness of the verification, perhaps using proof certificates (\eg{} à la~\cite{CMZ15}).

\ifdefined\VersionAuthor
	\subsection*{Acknowledgments}
\else
	\begin{acks}
\fi
We warmly thank the reviewers for their numerous useful suggestions.
We also thank the developers of \AnimUML{}, \EMI{} and \hugort{} for their useful precisions concerning their tools.
	This work is supported by project 9.10.11 %
		``Software Verification from Design to Implementation'' of French-Singaporean Programme Merlion. %
	We would also like to thank
		Mohamed Mahdi Benmoussa for his help with \cref{fig:SMD}.
\ifdefined\VersionAuthor
\else
	\end{acks}
\fi

\AuthorVersion{%
	\newpage{}
}

\ifdefined\VersionAuthor
	\newcommand{\CCIS}{Communications in Computer and Information Science}
	\newcommand{\ENTCS}{Electronic Notes in Theoretical Computer Science}
	\newcommand{\FAC}{Formal Aspects of Computing}
	\newcommand{\FundInf}{Fundamenta Informaticae}
	\newcommand{\FMSD}{Formal Methods in System Design}
	\newcommand{\IJFCS}{International Journal of Foundations of Computer Science}
	\newcommand{\IJSSE}{International Journal of Secure Software Engineering}
	\newcommand{\IPL}{Information Processing Letters}
	\newcommand{\JAIR}{Journal of Artificial Intelligence Research}
	\newcommand{\JLAP}{Journal of Logic and Algebraic Programming}
	\newcommand{\JLAMP}{Journal of Logical and Algebraic Methods in Programming} %
	\newcommand{\JLC}{Journal of Logic and Computation}
	\newcommand{\LMCS}{Logical Methods in Computer Science}
	\newcommand{\LNCS}{Lecture Notes in Computer Science}
	\newcommand{\RESS}{Reliability Engineering \& System Safety}
	\newcommand{\RTS}{Real-Time Systems}
	\newcommand{\SCP}{Science of Computer Programming}
	\newcommand{\SOSYM}{Software and Systems Modeling ({SoSyM})}
	\newcommand{\STTT}{International Journal on Software Tools for Technology Transfer}
	\newcommand{\TCS}{Theoretical Computer Science}
	\newcommand{\TOPLAS}{{ACM} Transactions on Programming Languages and Systems ({ToPLAS})}
	\newcommand{\ToPNoC}{Transactions on Petri Nets and Other Models of Concurrency}
	\newcommand{\TOSEM}{{ACM} Transactions on Software Engineering and Methodology ({ToSEM})}
	\newcommand{\TSE}{{IEEE} Transactions on Software Engineering}
\else
	\newcommand{\CCIS}{CCIS}
	\newcommand{\ENTCS}{ENTCS}
	\newcommand{\FAC}{FAC}
	\newcommand{\FundInf}{FI}
	\newcommand{\FMSD}{FMSD}
	\newcommand{\IJFCS}{IJFCS}
	\newcommand{\IJSSE}{IJSSE}
	\newcommand{\IPL}{IPL}
	\newcommand{\JAIR}{JAIR}
	\newcommand{\JLAP}{JLAP}
	\newcommand{\JLAMP}{JLAMP} %
	\newcommand{\JLC}{JLC}
	\newcommand{\LMCS}{LMCS}
	\newcommand{\LNCS}{LNCS}
	\newcommand{\RESS}{RESS}
	\newcommand{\RTS}{RTS}
	\newcommand{\SCP}{SCP}
	\newcommand{\SOSYM}{{SoSyM}}
	\newcommand{\STTT}{STTT}
	\newcommand{\TCS}{TCS}
	\newcommand{\TOPLAS}{ToPLAS}
	\newcommand{\ToPNoC}{ToPNOC}
	\newcommand{\TOSEM}{ToSEM}
	\newcommand{\TSE}{TSE}
\fi
\small
\ifdefined\VersionAuthor
	\renewcommand*{\bibfont}{\small}
	\printbibliography[title={References}]
\else
	\bibliographystyle{ACM-Reference-Format}
	\bibliography{literatureReview}                                           %
\fi

\end{document}